\documentclass[twocolumn]{aastex62}

\usepackage{longtable}
\usepackage[flushleft]{threeparttable}
\usepackage{tabularx}
\usepackage{multirow}
\usepackage{graphicx}
\usepackage{amsmath,amssymb}
\usepackage{color}
\usepackage{units}
\usepackage{epstopdf}
\usepackage{hyperref}
\usepackage{multirow}
\usepackage{url}
\usepackage{subfigure}
\usepackage{rotating}
\usepackage{enumitem}\setlist[description]{font=\textendash\enskip\scshape\bfseries}
\usepackage{lineno}
\newcommand{\beq}{\begin{equation}}
\newcommand{\eeq}{\end{equation}}
\newcommand{\bdm}{\begin{displaymath}}
\newcommand{\edm}{\end{displaymath}}

\definecolor{Gray}{gray}{0.9}
\definecolor{orange}{rgb}{0.9,0.5,0}

\begin{document}

\title{Fast-transient Searches in Real Time with ZTFReST: Identification of 
Three Optically-discovered Gamma-ray Burst Afterglows and New Constraints on the Kilonova Rate}

\author[0000-0003-3768-7515]{Igor Andreoni}
\affil{Division of Physics, Mathematics and Astronomy, California Institute of Technology, Pasadena, CA 91125, USA}

\author[0000-0002-8262-2924]{Michael W. Coughlin}
\affil{School of Physics and Astronomy, University of Minnesota, Minneapolis, Minnesota 55455, USA}

\collaboration{these authors contributed equally to this work}

\author[0000-0002-7252-3877]{Erik C. Kool}
\affiliation{The Oskar Klein Centre, Department of Astronomy, Stockholm University, AlbaNova, SE-106 91 Stockholm, Sweden}

\author{Mansi M. Kasliwal}
\affiliation{Division of Physics, Mathematics and Astronomy, California Institute of Technology, Pasadena, CA 91125, USA}

\author{Harsh Kumar}
\affiliation{Indian Institute of Technology Bombay, Powai, Mumbai 400076, India}

\author[0000-0002-6112-7609]{Varun Bhalerao}
\affiliation{Indian Institute of Technology Bombay, Powai, Mumbai 400076, India}

\author[0000-0002-3498-2167]{Ana Sagu{\'e}s Carracedo}
\affiliation{The Oskar Klein Centre, Department of Physics, Stockholm University, AlbaNova, SE-106 91 Stockholm, Sweden}

\author{Anna Y.~Q. Ho}
\affiliation{Department of Astronomy, University of California, Berkeley, CA 94720-3411, USA}
\affiliation{Lawrence Berkeley National Laboratory, 1 Cyclotron Road, MS 50B-4206, Berkeley, CA 94720, USA}
\affiliation{Miller Institute for Basic Research in Science, 468 Donner Lab, Berkeley, CA 94720, USA}

\author{Peter T.~H.~Pang}
\affil{Nikhef, 1098 XG Amsterdam, The Netherlands}
\affil{Department of Physics, Utrecht University, 3584 CC Utrecht, The Netherlands}

\author{Divita Saraogi}
\affiliation{Indian Institute of Technology Bombay, Powai, Mumbai 400076, India}

\author{Kritti Sharma}
\affiliation{Indian Institute of Technology Bombay, Powai, Mumbai 400076, India}

\author{Vedant Shenoy}
\affiliation{Indian Institute of Technology Bombay, Powai, Mumbai 400076, India}

\author{Eric Burns}
\affiliation{Department of Physics and Astronomy, Louisiana State University, Baton Rouge, LA 70803, USA}

\author[0000-0002-2184-6430]{Tom{\'a}s Ahumada}
\affiliation{Department of Astronomy, University of Maryland, College Park, MD 20742, USA}

\author[0000-0003-3768-7515]{Shreya Anand}
\affil{Division of Physics, Mathematics and Astronomy, California Institute of Technology, Pasadena, CA 91125, USA}

\author[0000-0001-9898-5597]{Leo P. Singer}
\affiliation{Astrophysics Science Division, NASA Goddard Space Flight Center, MC 661, Greenbelt, MD 20771, USA}
\affiliation{Joint Space-Science Institute, University of Maryland, College Park, MD 20742, USA}

\author{Daniel A. Perley}
\affiliation{Astrophysics Research Institute, Liverpool John Moores University, \\ IC2, Liverpool Science Park, 146 Brownlow Hill, Liverpool L3 5RF, UK}

\author{Kishalay De}
\affil{Division of Physics, Mathematics and Astronomy, California Institute of Technology, Pasadena, CA 91125, USA}

\author{U.C. Fremling}
\affil{Division of Physics, Mathematics and Astronomy, California Institute of Technology, Pasadena, CA 91125, USA}

%%%%%%%%%%%%%%%%%%%
% Alphabetical

\author[0000-0001-8018-5348]{Eric C. Bellm}
\affiliation{DIRAC Institute, Department of Astronomy, University of Washington, 3910 15th Avenue NE, Seattle, WA 98195, USA} 

\author{Mattia Bulla}
\affiliation{The Oskar Klein Centre, Department of Astronomy, Stockholm University, AlbaNova, SE-106 91 Stockholm, Sweden}

\author{Arien Crellin-Quick}
\affiliation{Department of Astronomy, University of California, Berkeley, CA 94720-3411, USA}

\author{Tim Dietrich}
\affiliation{Institut f\"{u}r Physik und Astronomie, Universit\"{a}t Potsdam, 14476 Potsdam, Germany}
\affiliation{Max Planck Institute for Gravitational Physics (Albert Einstein Institute), Am M\"uhlenberg 1, Potsdam 14476, Germany}

\author{Andrew Drake}
\affiliation{Division of Physics, Mathematics and Astronomy, California Institute of Technology, Pasadena, CA 91125, USA}

\author[0000-0001-5060-8733]{Dmitry A. Duev}
\affiliation{Division of Physics, Mathematics and Astronomy, California Institute of Technology, Pasadena, CA 91125, USA}

\author{Ariel Goobar}
\affil{The Oskar Klein Centre, Department of Physics, Stockholm University, AlbaNova, SE-106 91 Stockholm, Sweden}

\author{Matthew J. Graham}
\affiliation{Division of Physics, Mathematics and Astronomy, California Institute of Technology, Pasadena, CA 91125, USA}

\author[0000-0001-6295-2881]{David~L.\ Kaplan}
\affiliation{Center for Gravitation, Cosmology and Astrophysics, Department of Physics, University of Wisconsin--Milwaukee, P.O.\ Box 413, Milwaukee, WI 53201, USA}

\author{S. R. Kulkarni}
\affil{Division of Physics, Mathematics and Astronomy, California Institute of Technology, Pasadena, CA 91125, USA}

\author[0000-0003-2451-5482]{Russ R. Laher}
\affiliation{IPAC, California Institute of Technology, 1200 E. California Blvd, Pasadena, CA 91125, USA}

\author[0000-0003-2242-0244]{Ashish~A.~Mahabal}
\affiliation{Division of Physics, Mathematics and Astronomy, California Institute of Technology, Pasadena, CA 91125, USA}
\affiliation{Center for Data Driven Discovery, California Institute of Technology, Pasadena, CA 91125, USA}

\author[0000-0003-4401-0430]{David L. Shupe}
\affiliation{IPAC, California Institute of Technology, 1200 E. California Blvd, Pasadena, CA 91125, USA}

\author{Jesper Sollerman}
\affiliation{The Oskar Klein Centre, Department of Astronomy, Stockholm University, AlbaNova, SE-106 91 Stockholm, Sweden}

\author{Richard Walters}
\affiliation{Caltech Optical Observatories, California Institute of Technology, Pasadena, CA 91125, USA}

\author{Yuhan Yao}
\affil{Division of Physics, Mathematics and Astronomy, California Institute of Technology, Pasadena, CA 91125, USA}

\begin{abstract}

While optical surveys regularly discover slow transients like supernovae on their own, the most common way to discover extragalactic fast transients, which fade within a few nights in the optical, is via follow-up observations of gamma-ray burst and gravitational-wave triggers. However, wide-field surveys have the potential to also identify rapidly fading transients, including counterparts to multi-messenger sources, independently of such external triggers.
The volumetric survey speed of the Zwicky Transient Facility (ZTF), in particular, makes the survey sensitive to objects that are as faint and fast-fading as kilonovae, the optical counterparts to binary neutron stars and neutron star--black hole mergers, out to almost 200\,Mpc.
In this paper, we introduce an open-source software infrastructure, the ZTF REaltime Search and Triggering, \texttt{ZTFReST}, which is designed to identify kilonovae and fast optical transients in ZTF data. Using the ZTF alert stream combined with forced point spread function photometry, we have implemented automated candidate ranking based on their photometric evolution and fitting to kilonova models.
Automated triggering of follow-up systems, such as Las Cumbres Observatory, for sources that pass user-defined thresholds, has also been implemented. In 13 months of science validation, we found several extragalactic fast transients independent of any external trigger (though some counterparts were identified later), including at least one supernova with post-shock cooling emission (ZTF21aabxjqr), two known afterglows with an associated gamma-ray burst (ZTF20abbiixp, ZTF20abwysqy), two known afterglows without any known gamma-ray counterpart (ZTF20aajnksq, ZTF21aaeyldq), and three new fast-declining sources (ZTF20abtxwfx, ZTF20acozryr, and ZTF21aagwbjr) that are likely associated with GRB~200817A, GRB~201103B, and GRB~210204A. However, we have not found any objects which appear to be kilonovae; therefore, we constrain the rate of GW170817-like kilonovae to $R < 900$\,Gpc$^{-3}$\,yr$^{-1}$.  
A framework such as \texttt{ZTFReST} could become a prime tool for kilonova and fast transient discovery with the Vera C.~Rubin Observatory's Legacy Survey of Space and Time.
\end{abstract}

\section{Introduction}

Multi-messenger sources of astrophysical transients are changing time-domain astronomy. With a variety of survey facilities now online, there are numerous examples of systems making detections of these sources in the optical possible. These include the Panoramic Survey Telescope and Rapid Response System (Pan-STARRS; \citealt{MoKa2012}), Asteroid Terrestrial-impact Last Alert System (ATLAS; \citealt{ToDe2018}), the Dark Energy Camera (DECam; \citealt{FlDi2015}), the Zwicky Transient Facility (ZTF; \citealt{Bellm:19:ZTFScheduler,Graham2018,Masci2019,DeSm2018}), and, in the near future, BlackGEM \citep{BlGr2015} and the Vera C.~Rubin Observatory's Legacy Survey of Space and Time (LSST; \citealt{Ivezic2019}).

Relevant for optical fast-transient discovery have been searches for afterglows from gamma--ray bursts (GRBs; \citealt{KlSt1973,MeDj1997,GeMe2012}); these include both ``short'' and ``long'' classes \citep{Kouveliotou1993}, although this classification is subject to debate \citep{NoBo2006,ZhCh2008,BrNa2013,BuCo2016}. These sources have been identified by GRB survey instruments such as the {\it Neil Gehrels Swift Observatory} mission \citep{GeCh2004} and the Gamma--ray Burst Monitor (GBM; \citealt{MeLi2009}) onboard the {\it Fermi} satellite. In addition to many afterglow detections associated with Swift, dedicated follow-up of GBM sources in particular by both the Palomar Transient Factory (PTF; \citealt{LaKu2009}) and ZTF at Palomar Observatory have yielded afterglow detections as well \citep{SiKa2015,CoAh2019,AhEA2020}.

During LIGO and Virgo's second observing run, the detection of GW170817~\citep{AbEA2017b}, its burst of gamma-rays GRB~170817A \citep{GoVe2017,SaFe2017,AbEA2017c}, its short GRB afterglow \citep{2017ApJ...848L..21A,2017ApJ...848L..25H,2017Sci...358.1579H,2017ApJ...848L..20M,2017Natur.551...71T} and an optical/infrared ``kilonova" counterpart, AT2017gfo \citep{AnAc2017, ArHo2017, HuWu2017, ChBe2017,2017Sci...358.1556C,CoBe2017,2017Sci...358.1570D,DiMa2017,2017Sci...358.1565E,KaNa2017,KiFo2017,LiGo2017,2017ApJ...848L..32M,NiBe2017,2017Sci...358.1574S,2017Natur.551...67P,SmCh2017,TaLe2017, 2017PASJ...69..101U}, introduced the world to the science of counterparts to gravitational waves (GW) detected by Advanced LIGO \citep{aLIGO}, Advanced Virgo \citep{adVirgo}, and in the future, KAGRA \citep{Som2012}. The detection and characterization of kilonovae enable constraints on the neutron star equation of state \citep{BaJu2017, MaMe2017, CoDi2018b, CoDi2018, CoDi2019b, AnEe2018, MoWe2018,RaPe2018,Lai2019,DiCo2020}, the Hubble constant \citep{CoDi2019,CoAn2020,2017Natur.551...85A,HoNa2018,DiCo2020}, and r-process nucleosynthesis \citep{ChBe2017,2017Sci...358.1556C, CoBe2017,PiDa2017,RoFe2017,SmCh2017,WaHa2019,KaKa2019}. 

With the end of the third LIGO-Virgo observing run (O3), and with the entrance of KAGRA, without a viable counterpart to a binary neutron star or neutron star--black hole merger candidate \citep[e.g.,][]{Andreoni2019S190510g, CoAh2019b, Goldstein2019S190426c, Gomez2019, LuPa2019, AnCo2020, Ackley2020, Andreoni2020S190814bv, AnAg2020,GoCu2020,KaAn2020}, it becomes particularly urgent to continue the search for such objects in optical, wide-field survey data, independently of other multi-messenger and multi-wavelength triggers. 
These searches also serve as unbiased surveys for optical emission, with the potential to discover, for example, collapsars with dirty fireballs \citep{Dermer2000} that do not have prompt GRB emission \citep{Dermer2000, Huang2002, Rhoads2003}, or study whether optically identified kilonovae differ from those identified with gravitational-wave detections, while also enabling many of the studies of both cosmology and nuclear physics identified above.
In this work, we will refer to ``serendipitous'' observations (and discoveries) as those performed within routine survey observations, as opposed to ``triggered'' target of opportunity (ToO) observations, which use timing and/or localization information from other wavelengths or messengers. 

There are several differences between serendipitous and ToO searches. When a GW, GRB, or neutrino alert is issued, it is possible to perform dedicated, high-cadence ToO observations of these fields, using either a synoptic or a galaxy-targeted strategy \citep{GeCa2016}. In addition to localization information, a trigger also provides an explosion time to which we can compare all of the transients in the alert stream.
Serendipitous observations, on the other hand, do not rely on another detector to have found an astrophysical transient first, and therefore have neither localization nor explosion time information. For this reason, rare fast transients can be more difficult to pick out. Survey data also provide us with a much larger number of images to mine, which is technically challenging, but at the same time could offer a broad range of discovery opportunities.
Serendipitous searches of this type have already been successful in the cases of GRB afterglows \citep{CeUr2015,StTo2017,HoKu2018,AnKo2020,KaAn2020} and afterglows with no GRB detected \citep{CeKu2013,HoPe2020}. Multi-facility programs such as the ``Deeper, Wider, Faster" program \citep[][Cooke et al., in preparation]{Andreoni2020MNRAS} aim discovering counterparts to fast radio bursts and other elusive transients via simultaneous multi-facility observations at many different wavelengths.  

We are motivated to search for serendipitous kilonovae in optical survey data.
Unfortunately, kilonovae and gamma-ray burst afterglows, the objective of this study, rapidly fade in the optical (on timescales of a night), and therefore are more difficult to detect than other transients such as supernovae, often identifiable for weeks to months. In addition, kilonovae are expected to rapidly redden with time, making their identification potentially easier but detection possibly harder in optical bands.

There are many more transients from the alert generators than can be characterized in these modes of operation, due to limited follow-up telescope time. For example, ZTF can generate more than a million alert packets per night \citep{Patterson2018}, thus providing the transient community with a preview of the experience expected for the LSST data stream. However, as of December 2020, only $\sim$\,10\% of transients reported on the Transient Name Server (TNS) have been spectroscopically classified \citep{Kul2020}, predominantly due to lack of observation time. As kilonovae are inherently faint, it is relatively unlikely for them to be classified in routine spectroscopic follow-up of bright transients such as through the Bright Transient Survey \citep{FrMi2019}. Galaxy-targeted searches such as the Census of the Local Universe program \citep{De2020} are sensitive to dimmer sources, although the project relies only on ZTF alerts and is limited by galaxy catalog completeness.

In \cite{AnKo2020}, we presented kilonova rate constraints from archival searches of serendipitous observations from March 2018 to February 2020. These observations were during ZTF Phase I, which covered March 2018 to September 2020; ZTF Phase II has been ongoing since then. In \cite{AnKo2020}, several candidates were identified where real time follow-up would have greatly improved our ability to confirm the nature of the fast transients.
This fact motivated us to automate discovery and follow-up infrastructure to rapidly identify fast transients in optical survey data; other surveys such as Pan-STARRS are also undertaking dedicated searches of this kind \citep{McSm2020}.
This infrastructure is inspired by brokers such as the Alert Management, Photometry and Evaluation of Lightcurves (AMPEL; \citealt{Nordin:2019kxt, 2018PASP..130g5002S}) and builds upon existing tools such as the ``Target and Observation Managers'' (TOMs) being built by Las Cumbres Observatory and others \citep{StBo2018}, or the automatic triggering capabilities already implemented in AMPEL \citep{Nordin:2019kxt}.

We have developed automated filtering and follow-up infrastructure designed to perform a serendipitous search for kilonovae and afterglows known as ZTF Realtime Search and Triggering, \texttt{ZTFReST}.
In this paper, we describe the \texttt{ZTFReST} automated infrastructure and first results obtained during science validation.
We describe the algorithms and their implementation in \S\ref{sec:ztfrest}.
The science validation and early results are detailed in \S\ref{sec:science}. Three case studies are extensively presented in \S\ref{sec:case studies}, to demonstrate the type of multi-wavelength analysis made possible when fast transients are found serendipitously in the survey. We translate our non-detection of kilonovae into rate limits in Section~\ref{sec:kilonova_rate}. We summarize our conclusions and future outlook in \S\ref{sec:conclusion}.

\section{\texorpdfstring{ZTFR\MakeLowercase{e}ST}{ZTFReST}}
\label{sec:ztfrest}

\begin{figure*}[t]
\begin{center}
 \includegraphics[width=6.5in]{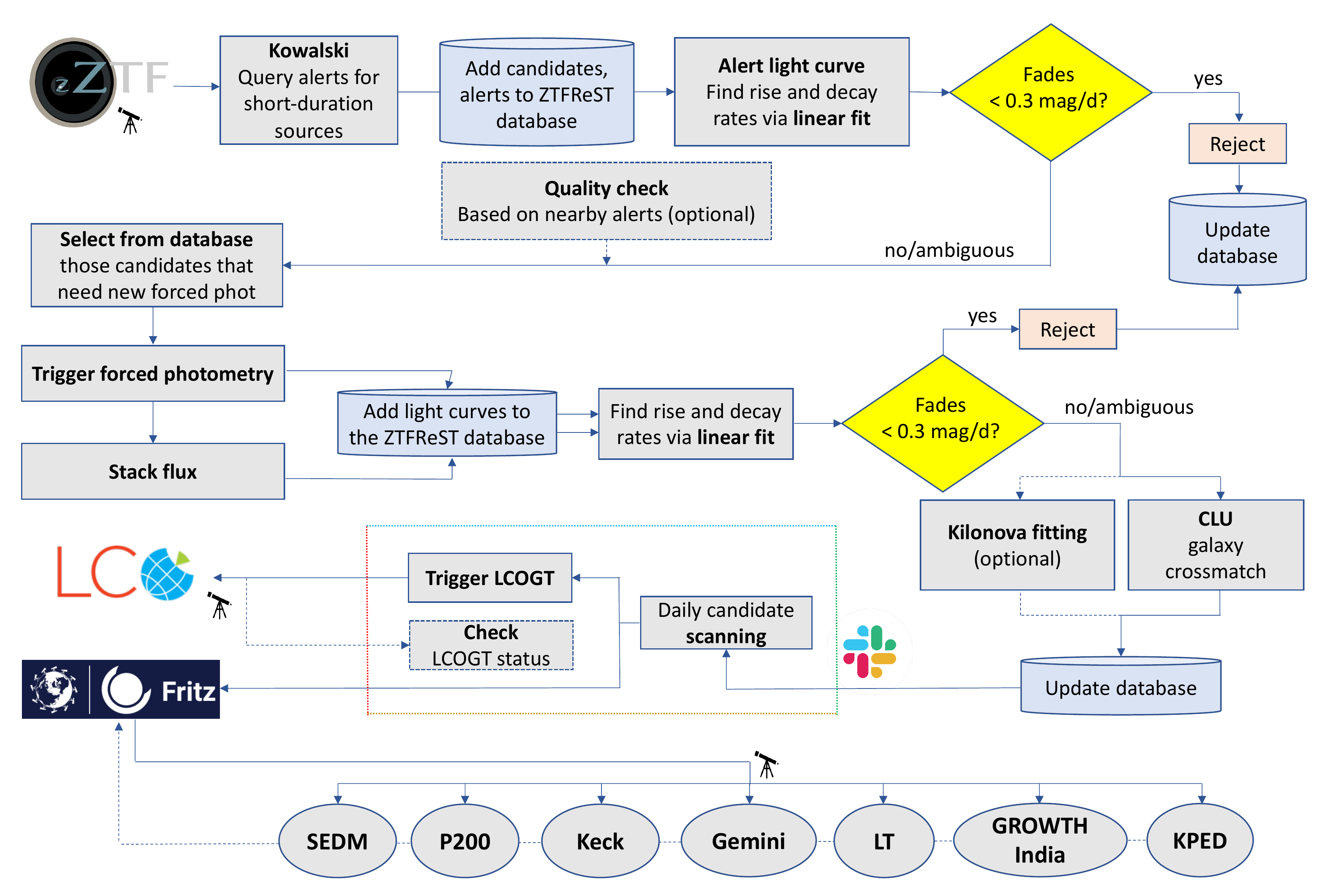}
  \caption{\texttt{ZTFReST} flowchart. Alerts are queried using \texttt{Kowalski} and their light curves undergo a first selection, where slow-evolving and long-duration transients are rejected. Then, a second selection is performed on PSF forced photometry and nightly-binned ``stacked" photometry. During daily scanning using the \texttt{Slack} application, transients are prioritized based on their fade rate and possible association with nearby galaxies present in the CLU catalog. Follow-up photometry with Las Cumbres Observatory (LCO) telescopes is automatically triggered for kilonova candidates directly from \texttt{Slack}. Finally, LCO data are downloaded and processed with an external image-subtraction and forced photometry pipeline. The most interesting candidates, along with LCO photometry results, are then uploaded to the ZTF Phase II marshal, known as Fritz.}
 \label{fig:flowchart}
\end{center}
\end{figure*}

A solid identification of rare transients such as kilonovae and orphan afterglows require multi-band, and often, multi-wavelength or multi-messenger data. To make this possible, one must first determine which of the many transients identified could be objects of interest based on their magnitude and color evolution, amongst other parameters. Once a strong candidate has been identified, the goal is to perform spectroscopic classification, when possible, and build a well-sampled, multi-wavelength light curve to characterize the system. This is particularly interesting for kilonovae as photometry and spectroscopy make it possible to extract information about the ejecta and therefore the original progenitor system \citep[e.g.,][]{CoDi2018}. Given the overall interest in kilonovae, coupled with their observational properties, it is important to prioritize rapidly fading and/or reddening candidates with no history of variability. 

\texttt{ZTFReST} relies on the ZTF alert stream \citep{Patterson2018}, which reports information about all 5-$\sigma$ detections, including its magnitude, proximity to other sources and its previous history of detections, among other metrics; it uses the alert stream from both public and private surveys \citep{Bellm:19:ZTFScheduler}. The largest public survey during ZTF Phase I had a three-night cadence, while ZTF Phase II predominantly has a two-night cadence. The public surveys in both ZTF Phase I and II obtain one 30-second exposure in both $g$-band and $r$-band every night a field was observed. The largest private program has been an extragalactic transient survey with a one-day cadence, where $g$- and $r$-band exposures are obtained six times per night; this survey covers $\approx$\,3,000 square degrees at high Galactic latitude ($\sim$\,$|b_{\text{Gal}}| > 30$\,deg).

The flowchart in Figure~\ref{fig:flowchart} offers a visual summary of \texttt{ZTFReST}, which is publicly developed on GitHub\footnote{\url{https://github.com/growth-astro/ztfrest}}.
Much of the infrastructure is built upon the pipeline described in \cite{AnKo2020} for historical, serendipitous kilonovae searches. Here we describe the set of procedures that streamline the pipeline in order to perform near real-time searches.

\subsection{Alert stream database queries}
\label{subsec:kowalski}

The alerts from both public and private surveys are queried using a local instance of \texttt{Kowalski}\footnote{\url{https://github.com/dmitryduev/kowalski}}, an open-source, multi-survey data archive and alert broker \citep{DuMa2019}.
In addition to storing all ZTF alert/light curve data in a single MongoDB, \texttt{Kowalski} also has built-in capabilities of matching against external catalogs. We regularly query \texttt{Kowalski} to identify the transients that pass specific criteria. In particular, we are interested in transients that are i) astrophysical in nature, i.e. unlikely instrumental artifacts; ii) short-lived; iii) without previous history of variability; and iv) without a spatially coincident stellar counterpart. Appropriate queries to \texttt{Kowalski} can address these requirements. The real/bogus selection, which is used to ensure transients are astrophysical in nature, is based on a deep learning classification algorithm \citep{DuMa2019}. We employ a threshold of $\text{\texttt{drb}} > 0.9$, for which \cite{DuMa2019} measured a false positive rate of 0.4\% at the cost of a false negative rate of 5\%. For further details regarding the other query parameters, we refer the reader to \cite{AnKo2020}.

\subsection{Light curve generation and fitting}
\label{subsec:lightcurves}

The main focus of this pipeline is to find transients undergoing rapid luminosity evolution and, in particular, those that are rapidly fading. Therefore, we generate a set of light curves for each transient and fit them to a linear model in magnitude space, to measure either the decay or rise rates, in units of magnitudes per day. 
This metric has been used in the past \citep{KaAn2020} to model the luminosity evolution of kilonovae, and provides a straightforward approximation for the luminosity evolution over the timescales and passbands we are interested in here.
For each candidate, up to three types of light curves can be generated:
a light curve based on the content of the ZTF alerts, a PSF forced-photometry light curve, and a nightly ``stacked'' light curve, built by combining forced photometry measurements (see below).

The light curves from the alerts are compiled with the full candidate detection history (the \texttt{prv} field in the alert packets), thus including $3 \leq \text{S/N} \leq 5$ detections found by the ZTF pipeline prior to the initial $>$ 5\,$\sigma$ alert. First, we make a cut on the duration, both filter-agnostic and per-band. We select only candidates with a maximum total duration of 14 days (i.e. reject any candidates whose difference between first and last detection exceeds 14 days) and a maximum single-band duration of 10\,days, 12\,days, and 14\,days in $g$-, $r$-, and $i$-band respectively; these numbers were broadly tailored to a conservative estimate of kilonova fade rates and expected ZTF limits for kilonovae in the local universe.
Second, we perform linear fits in magnitude vs time space when light curves have multiple detections over at least a 0.5\,day baseline in a given band; to be as simple as possible, the fits are not weighted and no chi-squared metric or similar is evaluated.
We place a hard constraint of fading at least 0.3\,mag\,day$^{-1}$ in any one of the $g$-, $r$- or $i$-bands, as shown to be appropriate for a wide-range of kilonova model grids \citep{AnKo2020, KaAn2020}. Those candidates that do not pass this threshold are rejected from further consideration.

For those objects which are either fading faster than this threshold or for which there is not currently enough signal to noise in the existing observations to tell, we perform a data quality check based on nearby alerts. If there were recent alerts within $3''$ of the object and different ID (i.e., further than $1''$), it is possible that the candidates are artifacts caused by electronic crosstalk or by reflections within the optical system (known as ``ghosts''). For those objects not rejected by this criteria, PSF photometry is performed pinned to the median location of each candidates' set of alerts using \texttt{ForcePhotZTF} \citep{Yao2019}, on images processed with the ZTF pipeline at IPAC \citep{Masci2019} using the ZOGY image-subtraction algorithm \citep{ZaOf2016}. The precise coordinates of the candidates are obtained from the median location reported in the ZTF alerts.

At this point, we created ``stacked'' light curves, where the forced-photometry flux within each night in each band is ``optimally'' combined. Specifically, we use a weighted average when all the data points have $\text{S/N} \geq 3$, where the squares of the S/N are used as weights. If a night includes at least one data point with $\text{S/N} < 3$, a simple mean is used. Numerous tests revealed that stacking flux this way reduces significantly the number of spurious low-S/N detections, while providing us with deeper and more precise photometry of sources with $\text{S/N} < 3$ in individual exposures \citep[see Figure\,2 in][]{AnKo2020}. 

The fit to linear models is repeated for both the forced and stacked photometry. Once again, candidates failing the 0.3 mag\,day$^{-1}$ cut are rejected from further consideration if their light curves have sufficient signal to noise and baseline to fit to linear models.

\subsection{Galaxy catalog cross-matching}
\label{subsec:CLU}

Those objects that pass this step are cross-matched with galaxies further than 10\,Mpc in the Census of the Local Universe (CLU) catalog \citep{CoKa2017}, with a ``match'' declared if the catalog-reported location of the galaxy is within 100\,kpc of the transient's location \citep{Ber2014}; this catalog is especially useful given its completeness (85\% in star-formation and 70\% in stellar mass at 200\,Mpc). Catalogs like CLU and GLADE \citep{DaGa2018} have proven to be very useful for galaxy-targeted follow-ups of gravitational-wave events, including GW170817 \citep{ArHo2017,2017Sci...358.1556C,VaSa2017}, especially powerful given it reduces the sky area requiring covering to $\approx 1$\,\% within these local volumes \citep{CoKa2017}.
Given the intrinsically faint absolute magnitudes of kilonovae, the presence of a transient in a known nearby galaxy could make it particularly interesting.
Even if there is no match, we do include the candidate in the scanning step, i.e. this is a value-added diagnostic.

\subsection{Kilonova model fitting}
\label{subsec:modelfit}

At this point, we have two optional features implemented.
The first employs automated fits to kilonova light curve model grids, such as those provided by ~\cite{KaMe2017} and \cite{Bul2019}, combined with a Gaussian Process Regression framework~\citep{CoDi2018,CoDi2018b,DiCo2020}.
In particular, due to the limited number of light curve points, we reduce dimensionality with single component light curve models by default; however, multiple component models with both ``dynamical ejecta'' and disk winds driven by neutrino energy, magnetic fields, viscous evolution and/or 
nuclear recombination are also available \citep[e.g.,][]{MePi2008,BaGo2013,DiUj2017,SiMe2017}.
Unlike fitting to GW170817 or similar, neither the explosion time nor the distance is fixed, and therefore in addition to the model parameters such as the ejecta masses, ejecta velocities, lanthanide fractions, inclination angles, amongst others, both the distance modulus and the explosion time must be fit for.
For now, we optionally perform the fits; in the future, we desire to use these fits to prioritize follow-up resources and use the fit efficacy to assign a probability of ``discovery''.

\subsection{Semi-automatic follow-up triggering}
\label{subsec:trigger}
The second feature is to automatically trigger follow-up observations based on the Las Cumbres Observatory network \citep{StBo2018}. The ZTF survey cadence may be insufficient, on its own, to fully characterize the fast decaying light curves of afterglows and kilonovae, especially when considering loss of time due to weather or bright moon phases. For this reason, automated infrastructure to trigger on particularly interesting candidates has been implemented.

The objects surviving the selection criteria are scanned by on-duty astronomers within dedicated \texttt{Slack} application channels built for this purpose. We use a set of scores, built based on the presence of rapid decay, proximity to a CLU galaxy, distance from the Galactic plane, and others in order to prioritize the candidates for scanning; we do not make any cuts on Ecliptic latitude. Within the \texttt{Slack} channel, textual information such as the coordinates, fade rates, CLU galaxy crossmatch, Galactic latitude, and expected extinction are listed. We also display both the discovery, reference, and difference images, as well as the photometric time-series plotted separately for the alert, forced photometry, and stacked photometry data streams. From within \texttt{Slack}, we can trigger Las Cumbres Observatory network \citep{StBo2018} observations directly with a simple command. These data obtained with LCO are reduced automatically using a dedicated pipeline \citep{FrSo2016} and uploaded in the GROWTH Marshal \citep{Kasliwal2019} during ZTF Phase I and in the Fritz marshal \citep{WaCr2019,DuMa2019} during ZTF Phase II; these Marshals are used for examining all relevant proprietary and external data-sets rapidly, enabling communication between collaboration members and triggering further followup observations of interesting objects.

\section{Science Validation and First Results}
\label{sec:science}

Since 2020 September 21, we have been running \texttt{ZTFReST} every day. 
We validated the output of the pipeline by running it first on 265 days of ZTF ``archival" data, from 2020 January 1 to 2020 September 20; from then on, new data was processed daily. 
On $\sim$\,7\,\% of nights, the dome is closed and there is no new data; a further $\sim$\,3\,\% of nights have poor conditions such that the magnitude limits are 2 magnitudes brighter than the median limits for the survey.
A summary of the confirmed extragalactic fast transients (along with one yet un-classified source) identified during science validation can be found in Table\,\ref{table: candidates}.

\subsection{Science validation I: archival data}
\label{subsec:SV archival}

The  \texttt{ZTFReST} code allows us to easily input a range of dates to search for candidates, with the default being the last 24 hours from the time when the pipeline starts running. 

\begin{table}[ht]
\centering
\label{table:filtering_ztf}
\begin{tabular}{lp{25mm}}
\hline \hline
Filtering criteria & \# of Candidates \\ \hline 
Science Validation & 15,555 \\% \hline
$|b_{Gal}| > 10$\,deg & 12,710\\ %\hline
At least one band: 0.3\,mag\,day$^{-1}$ & 309\\ %\hline
All bands: 0.3\,mag\,day$^{-1}$ & 177\\ %\hline
Kilonova-like fade rates & 40\\ %\hline
Not previously identified CVs & 12\\ %\hline
Likely afterglows & 4\\ 
Likely kilonovae & 0\\
\hline 
\end{tabular}
\caption{Filtering results for \texttt{ZTFReST} on 265 days of archival science validation data. We show the number of transients that pass each step, having applied that criteria over the remaining transients from the previous stage. The criteria are further described in Section \ref{subsec:SV archival}.}
\end{table}

\begin{figure}[t]
    \centering
    \includegraphics[width=0.5\textwidth]{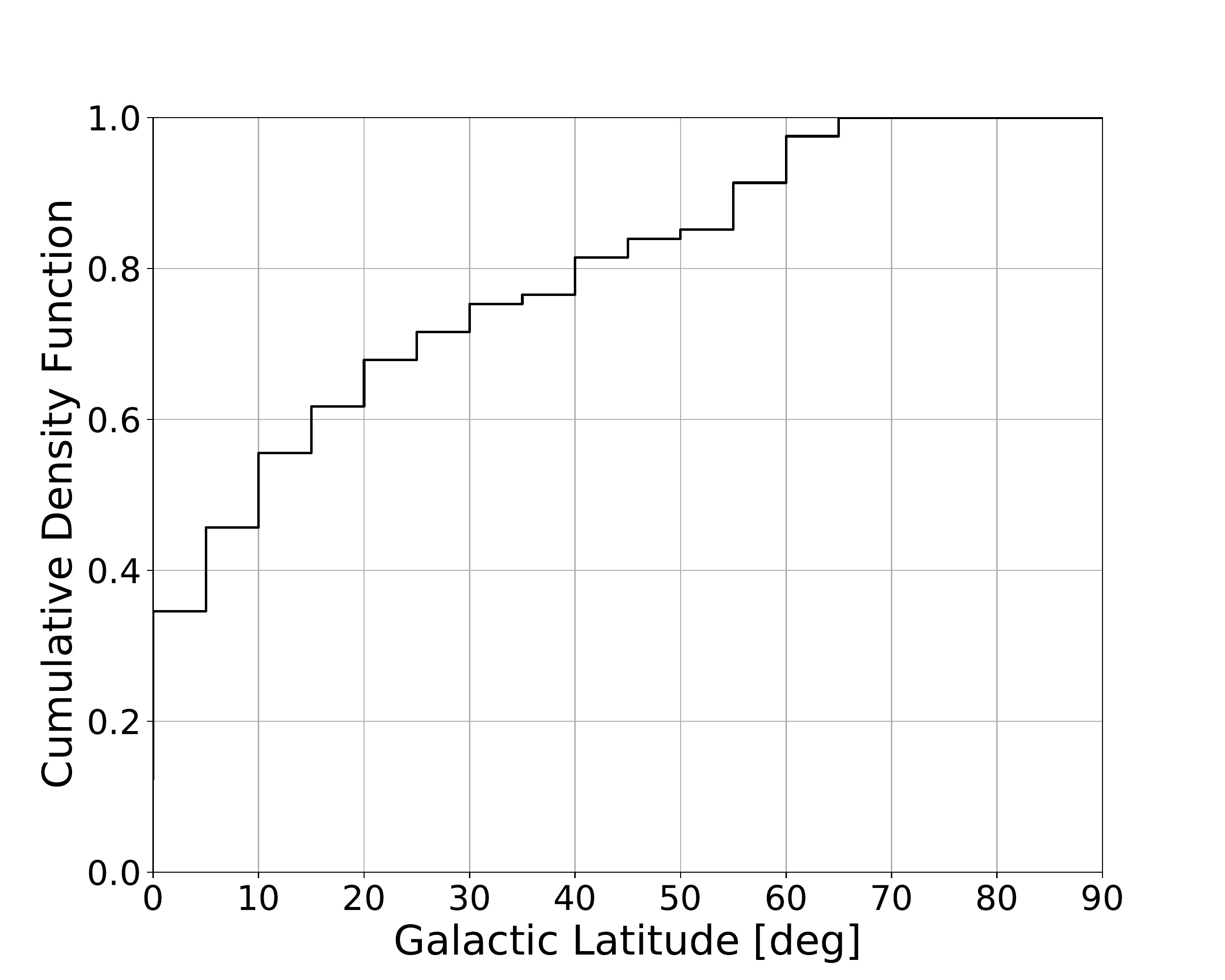}
    \caption{Cumulative density function of $|b_{Gal}|$ for those transients passing the criteria of fading faster than 0.3\,mag\,day$^{-1}$ in at least one band.}
    \label{fig:gal}
\end{figure}

The 265 days of data used for the science validation yielded 81,651,645 alerts in total. We identified %7,984
15,555
short-duration (at most 14 day) transient candidates using \texttt{Kowalski}. Since the main objective of this science validation is to understand the \texttt{ZTFReST} capability to find extragalactic fast transients, we limited our queries to high Galactic latitudes by imposing a $|b_{Gal}| > 10$\,deg cut, which brings the number of candidates to %4,363.
12,710.
The candidates with a fading rate faster than 0.3\,mag\,day$^{-1}$ in at least one band with at least one photometry method (alerts, forced photometry, or stacked forced photometry) are %347
309. Of these candidates, we reject 132, as they have a slow evolution (fade rate slower than 0.3\,mag\,day$^{-1}$) in at least one band; while this requirement is not used for our real-time processing (see below), it was useful to limit the number of science validation candidates.
This brings the number of candidates down to 177, of which %144
29
are located within 100\,kpc from CLU catalog galaxies further than 10\,Mpc.
To demonstrate the benefit of the Galactic latitude cut in particular, Figure~\ref{fig:gal} shows the cumulative density function of $|b_{Gal}|$ for those transients passing the criteria of fading faster than 0.3\,mag\,day$^{-1}$ in at least one band; unsurprisingly, $\sim$ 45\% of such transients are located $|b_{Gal}| \leq 10$; because there are so many stars near to the Galactic plane, the likelihood of identifying flaring stars there is very high, indicating the utility of such a cut to decrease the background of Galactic stars.

For science validation purposes, we applied criteria for light curve evolution based on kilonova models as in \citet{AnKo2020}. Thus, we selected those sources with fading rates larger (i.e., faster) than 0.57\,mag\,day$^{-1}$ in $g$-band, 0.39\,mag\,day$^{-1}$ in $r$-band, or 0.3\,mag\,day$^{-1}$ in $i$-band; again, this is a stricter cut than used in our real-time processing (see below).
Only 40 candidates passed the strict selection criteria. We vetted the candidates by inspecting, for each one of them, small cutouts of the science image, the reference image, and the image subtraction, the light curve built with information included in the ZTF alerts, the forced PSF photometry light curve, and the nightly stacked PSF photometry light curve.
Twelve candidates found during archival searches passed human inspection that were not already classified as cataclysmic variables (CVs). Of those, eight were excluded from the sample after further vetting. In particular, for one candidate, faint detections were revealed by forced photometry on every ZTF image available; three showed multiple outbursts in ATLAS, found using the forced photometry server \citep{ToDe2018, SmSm2020}; one had an underlying point source classified as ``PSF" by Legacy Survey DR9 \citep{DeSh2019} \texttt{tractor} modeling; one appears to be a very slow moving object.
Two are located at Galactic latitude $10 < |b_{\text{Gal}}| < 15$\,deg in crowded stellar fields, appear to be hostless, and show blue or ``grey" color ($g-r \leq 0$\,mag), which suggests that they are more likely stellar outbursts rather than cosmological afterglows, kilonovae, or other types of genuine extragalactic transients.

Four sources passed all of our tests and are either confirmed or likely GRB afterglows:

\begin{itemize}

\item ZTF20aajnksq (AT2020blt) was spectroscopically classified as an afterglow at redshift $z=2.9$, with no associated gamma--ray counterpart \citep{Ho2020}.

\item ZTF20abbiixp (AT2020kym) was likely the optical counterpart to GRB~200524A, already found serendipitously in ZTF data by \cite{2020GCN.27799....1H}. Optical spectroscopy suggests that the most probable redshift is $z=1.256$ \citep{GCN29673}. This object will be discussed in a paper in preparation on afterglows discovered in ZTF.

\item ZTF20abwysqy (AT2020scz) was the confirmed optical counterpart to the short-duration \\ GRB~200826A. This transient was found during rapid-response follow-up of a coarsely localized {\it Fermi} trigger with ZTF \citep{GCN28295}; it will be described in detail by Ahumada et al. (in preparation). Follow-up observations placed it at redshift $z=0.7481 \pm 0.0003$ \citep{gcn28319}.

\item ZTF20abtxwfx (AT2020sev) was not spectroscopically confirmed, but its fast-evolving light curve, the presence of a radio counterpart, and a possible association with GRB~200817A \citep{GCN_GRB200817A_GBM} suggest it being an afterglow as well \citep{2020ATel13978....1A}. The multi-wavelength analysis of this transient is presented in \S\ref{subsec:ZTF20abtxwfx}, and photometry is presented in Table\,\ref{tab:ZTF20abtxwfx}.

\end{itemize}
For a tabular summary of this discussion, please see Table\,\ref{table:filtering_ztf}.

In addition, we ran the pipeline on data from 2019, appropriately choosing time spans of $\pm 7$ days centered on the first detection of two known sources of interest. The former, ZTF19aabgebm (AT2019aacx), is the afterglow counterpart to GRB~190106A. The latter, ZTF19aanhtzz (AT2019aacu) is a transient found at high Galactic latitude ($b_{Gal} = 59$\,deg) that faded by $\Delta r = 1.24$\,mag\,day$^{-1}$ \citep{AnKo2020}. Both ZTF19aabgebm and ZTF19aanhtzz were successfully recovered.

\begin{table*}[]
%\begin{threeparttable}
\centering
    \begin{tabular}{llllrclcc}
   \hline \hline
Name & TNS & RA & Dec & $b_{\text{gal}}$  & Classification & References & Fade $g$ & Fade $r$ \\
& & &  & (deg) & or GRB & & (mag/d) & (mag/d) \\
\hline
%ZTF19aabgebm & AT2019aacx & 29.88002 & 23.84546 & --36.4 & afterglow & \cite{AnKo2020, Ho2020arXiv} \\
%ZTF19aanhtzz & AT2019aacu & 200.71300 & 57.46357 & 59.2 & afterglow? & \cite{AnKo2020} \\
\multirow{2}{*}{ZTF20aajnksq} & \multirow{2}{*}{AT2020blt} & \multirow{2}{*}{12:47:04.87} & \multirow{2}{*}{+45:12:02.25} & \multirow{2}{*}{71.9} & Afterglow & \multirow{2}{*}{[1]} & \multirow{2}{*}{-} & \multirow{2}{*}{1.58} \\
& & & & & no GRB & \\
ZTF20abbiixp & AT2020kym & 14:12:10.34 & +60:54:19.01 & 53.6 & GRB~200524A & [2] & - & 3.07 \\
ZTF20abwysqy & AT2020scz & 00:27:08.55 & +34:01:38.36 & --28.6 & GRB~200826A & [3] & 2.01 & - \\
{\bf ZTF20abtxwfx} & AT2020sev & 16:41:21.23 & +57:08:20.67 & 40.0 & GRB~200817A & this work; [4] & 0.49 & 0.48 \\
\hline
ZTF20acgigfo & AT2020urd & 00:40:31.10 & +40:35:53.90 & --22.2 & Nova & this work & - & 0.62 \\
ZTF20acstbfh & AT2020aapw & 00:40:19.74 & +40:49:35.82 & --22.0 & Nova & this work; [5] & 0.61 & - \\
{\bf ZTF20acozryr} & AT2020yxz & 02:48:44.33 & +12:08:14.16 & --41.5 & GRB~201103B & this work; [6] & 0.75 & 0.78 \\
ZTF21aaarlbp & AT2021bl & 01:33:21.99 & +30:33:01.27 & --31.5 & Nova & this work & 0.81 & 0.87 \\
ZTF21aabxjqr & SN2021pb & 09:44:46.80 & +51:41:14.41 & 47.4 & Shock cooling & this work; [7,8] & 0.40 & 0.28 \\
\multirow{2}{*}{ZTF21aaeyldq} & \multirow{2}{*}{AT2021any} & \multirow{2}{*}{08:15:15.34} & \multirow{2}{*}{--05:52:01.23} & \multirow{2}{*}{15.7} & Afterglow & \multirow{2}{*}{[9]} & \multirow{2}{*}{-}& \multirow{2}{*}{17.56} \\
& & & & & no GRB & & & \\
{\bf ZTF21aagwbjr} & AT2021buv & 07:48:19.30 & +11:24:34.32 & 17.7 & GRB~210204A & this work; [10] & - & 2.34 \\
ZTF21aahifke & AT2021clk & 02:54:27.54 & +36:31:56.74 & --20.1 & Unknown & this work; [11] & - & 0.96 \\
ZTF21aapkbav & AT2021gca & 14:28:07.33 & +33:29:49.38 & 68.2 & SN II & this work; [7,12] & 0.29 & 0.31\\
   \hline
    \end{tabular}
    \caption{Afterglows found with \texttt{ZTFReST} during science validation, in both archival searches (above the horizontal line; \S\ref{subsec:SV archival}) and in real-time (below the horizontal line; \S\ref{subsec:SV real time}). The names of the three new, confirmed afterglows discovered by \texttt{ZTFReST} are marked in boldface. For each transient, this table presents its Transient Name Server (TNS) denomination, coordinates (J2000), Galactic latitude, classification or associated GRB when known, discovery references, and fade rate at the time of discovery in $g$ and $r$ bands. In particular, the highest fade rates measured using ZTF alerts, forced photometry, and nightly stacked forced photometry are reported. References: [1] \cite{Ho2020}; [2] \cite{2020GCN.27799....1H}; [3] \cite{GCN28295}; [4] \cite{gcn28305}; [5] \cite{2020ATel14204....1T}; [6] \cite{CoAn2020b}; [7] Fremling et al., in prep; [8] \cite{2021ATel14320....1M} [9] \cite{gcn29305}; [10] \cite{gcn29405}; [11] \cite{gcn29446}; [12] \cite{astronote_ZTF21aapkbav_disc}.}
    \label{table: candidates}
%\end{threeparttable}
\end{table*}

\begin{figure*}[t]
    \centering
    \includegraphics[width=0.8\textwidth]{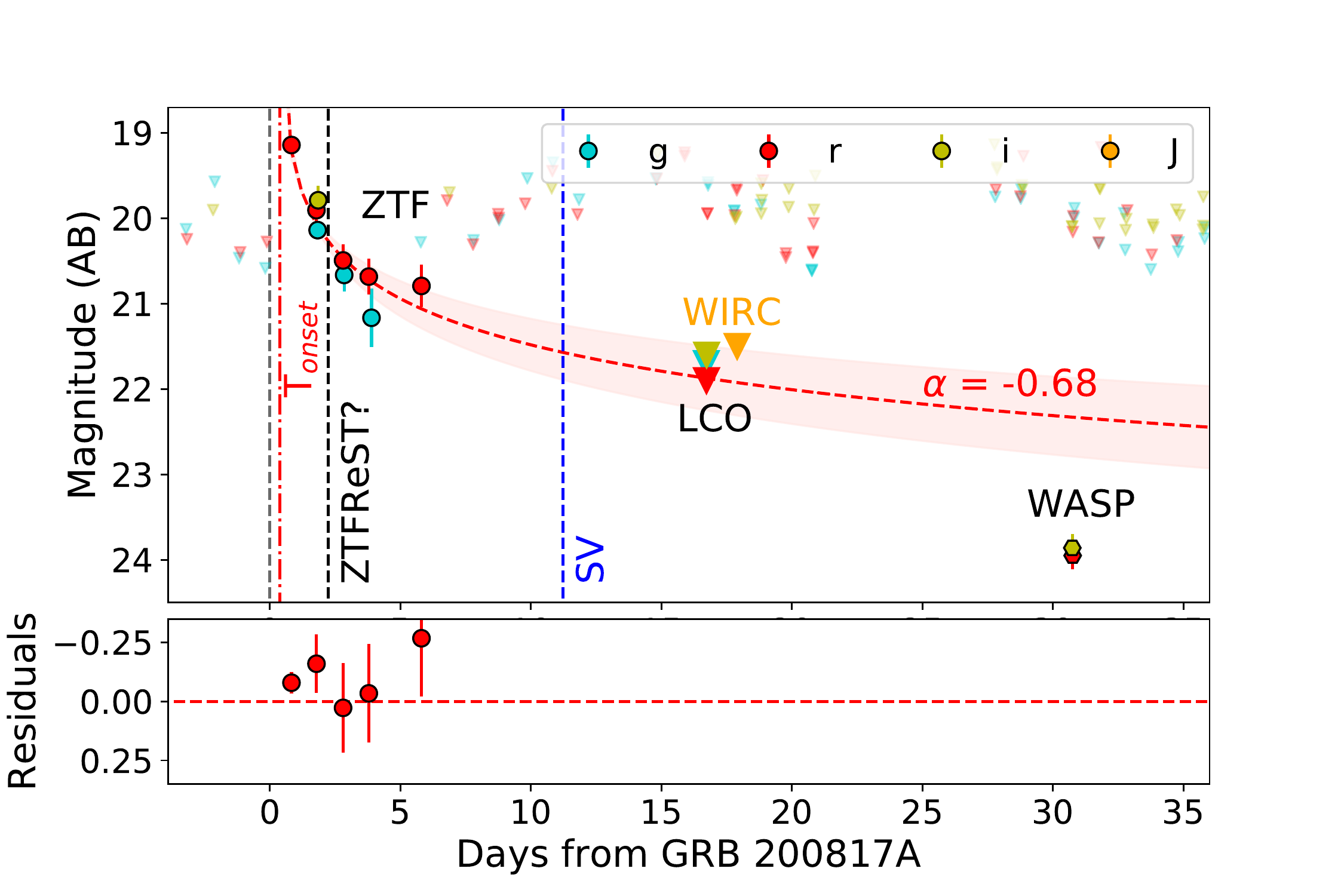}
    \includegraphics[width=0.8\textwidth]{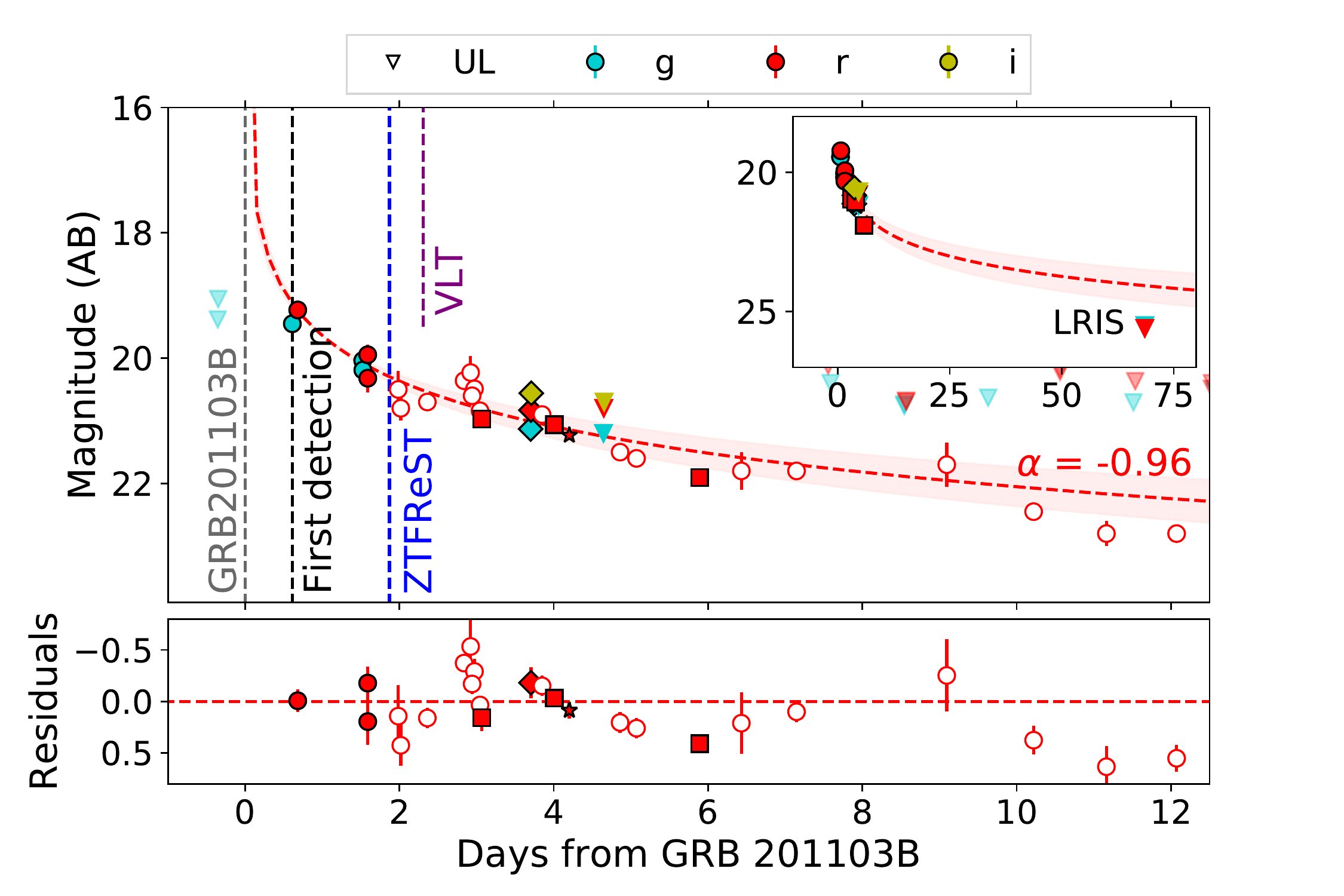}
    \caption{Forced PSF photometry light curve of the fast transients ZTF20abtxwfx (top panel) and ZTF20acozryr (bottom panel). {\it Top row}: In addition to ZTF optical data, the upper plot shows a $J$-band near-infrared upper limit (marked by triangles for all instruments) obtained with P200+WIRC and late-time optical imaging obtained with LCO and P200+WaSP (hexagons). The abscissa is centered on the detection time of GRB~200817A, under the assumption that it is the gamma--ray counterpart to ZTF20abtxwfx. The onset time estimated from the power-law fit, $T_{\text{onset}}$, is indicated by a vertical red line; the shaded region indicates the $1\sigma$ uncertainty on the power-law index; the fit residuals are shown in the bottom panel. The times of discovery during science validation (SV), the expected near real-time discovery time (during regular \texttt{ZTFReST} operations), and the GRB~201103B trigger time are indicated by vertical dashed lines. Fit residuals are shown for all $r$-band data points in the lower panel. {\it Bottom row}: In addition to ZTF optical data (solid circles), the upper plot shows our LCO (diamonds), LT (stars), and GIT (squares) follow-up observations. Open circles mark data points published by other groups via GCN circulars in $r$ or $R$ bands. The abscissa is centered on the detection time of GRB~201103B. A power-law fit was performed using only ZTF, LCO, LT and GIT $r$-band data for a fixed onset time, equal to the discovery time of GRB~201103B. Fit residuals are shown for all $r$-band data points in the lower panel. The near real-time discovery time of ZTF20acozryr is indicated with a blue dashed line. The purple dashed line marks the time when ZTF20acozryr was spectroscopically classified using X-Shooter on VLT \citep{XuVi2020}. }
    \label{fig:lightcurves}
\end{figure*}

\subsection{Science validation II: real time operations}
\label{subsec:SV real time}

Daily data processing with \texttt{ZTFReST} started on 2020 September 21. Every morning at 07:30\,AM Pacific Time, when all the images acquired during the night are processed by the image-subtraction pipeline at IPAC \citep{Masci2019}, a \texttt{crontab} automatically starts \texttt{ZTFReST}. After $\sim30$\,minutes, \texttt{ZTFReST} finishes all the operations and a bot announces to the team that the analysis is complete via the \texttt{Slack} application (see also \S\ref{subsec:trigger}). While this process is currently run once per night, we may explore running halfway through the night to potentially identify some fast transients earlier in the future, possible due to the participation by scientists worldwide.

In the first $\sim$\,4 months of operations, the pipeline has yielded between zero and eleven candidates per day by requiring a conservative fading rate larger than 0.3\,mag\,day$^{-1}$. The median and mean number of candidates to be scanned per day has been 2 and 2.5, respectively, with a standard deviation of 2.4 candidates per day. Similarly, there have been a mean of 0.5 new candidates per day, with a peak of 5. No cut on the Galactic latitude has been applied.

\begin{figure}[ht]
    \centering
    \includegraphics[width=\columnwidth]{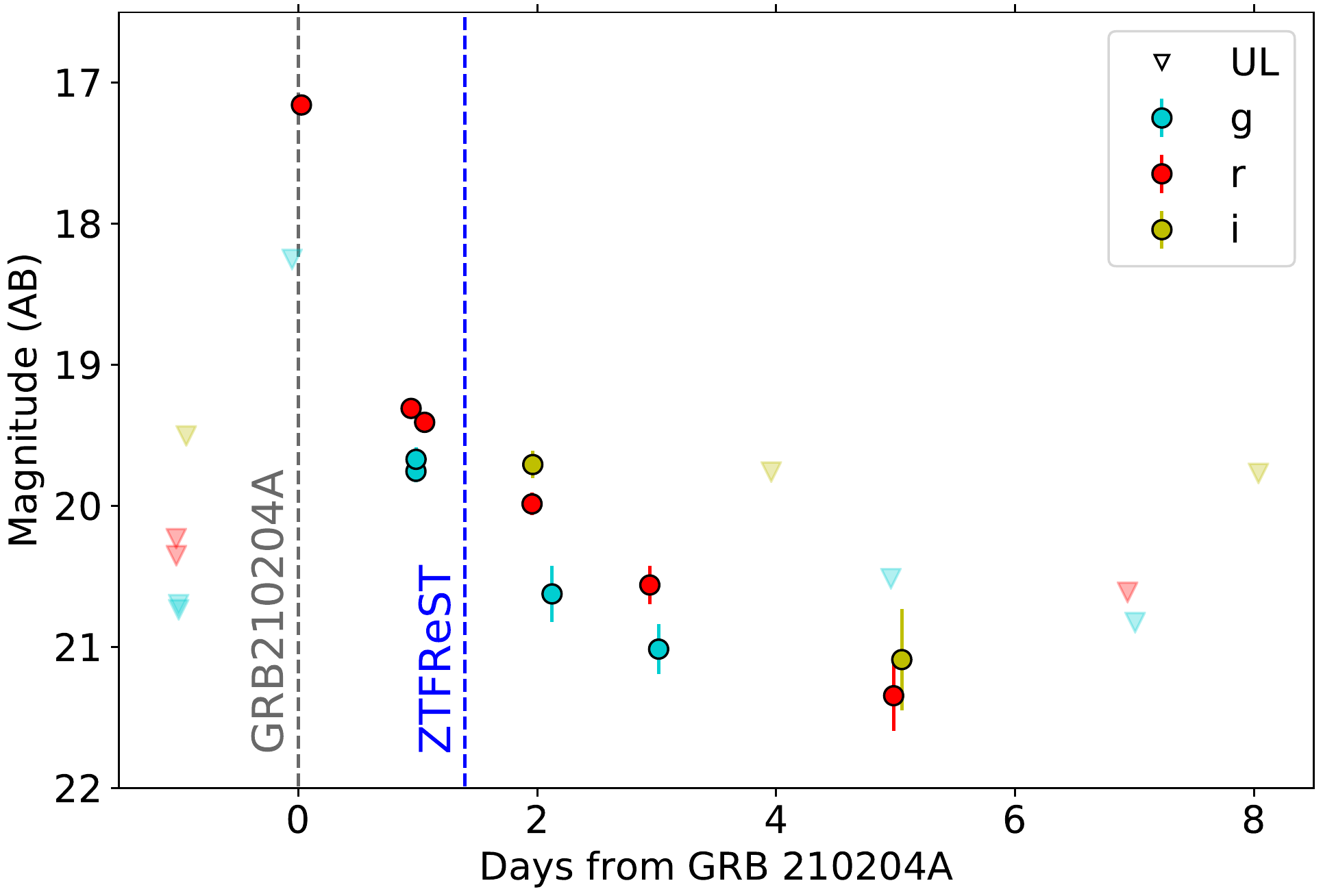}
    \includegraphics[width=\columnwidth]{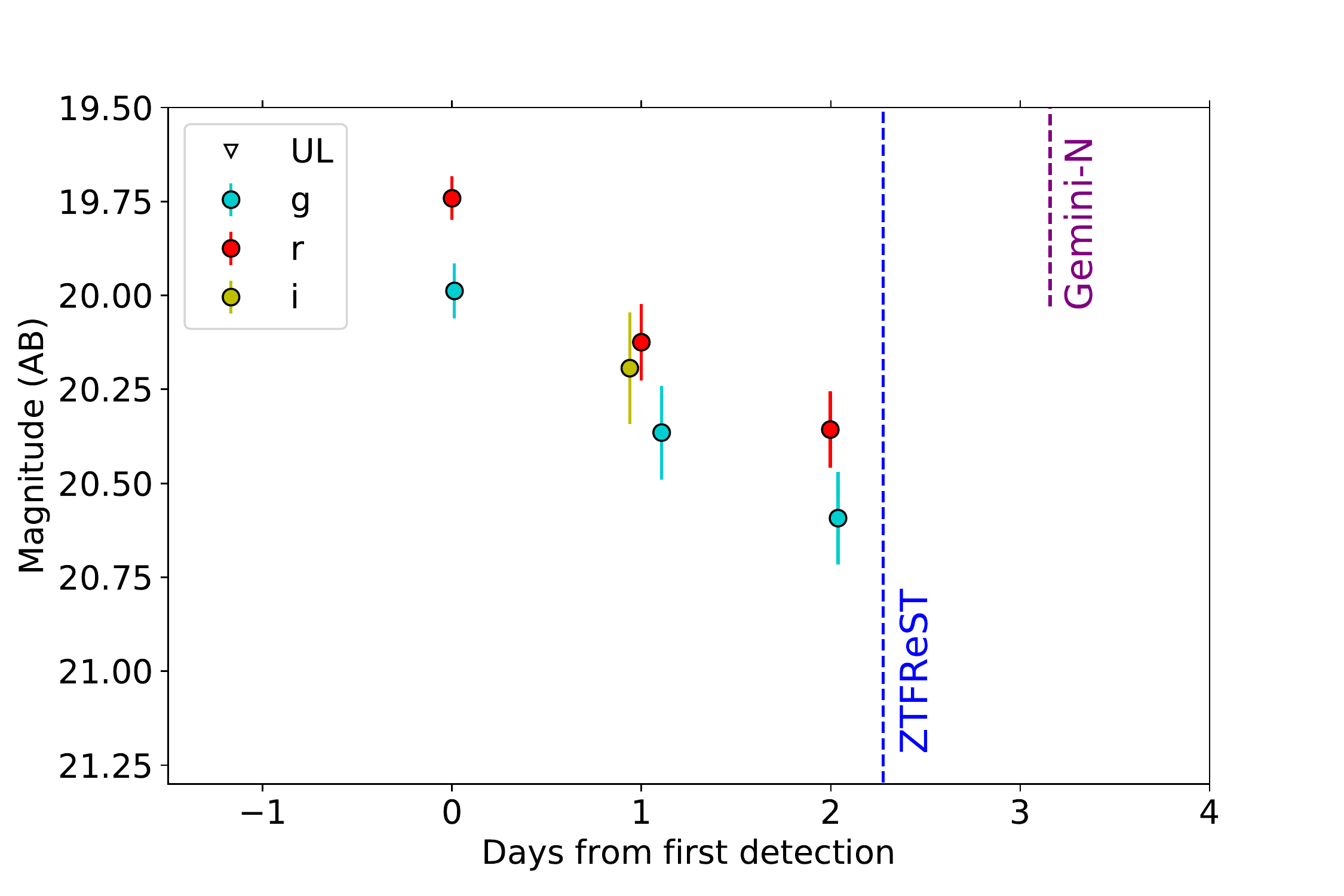}
    \caption{Photometric light curves of the fast transients ZTF21aagwbjr ({\it top}) and ZTF21aapkbav, the optically-discovered afterglow of GRB~210204A ({\it bottom}).}
    \label{fig:lightcurves bjr bav}
\end{figure}

Among the large number of sources found in near real-time, the transients we identified as extragalactic fast transients were:

\begin{itemize}

\item ZTF20acgigfo (AT2020urd) and ZTF20acstbfh (AT2020aapw) -- Novae in the M31 galaxy.

\item ZTF20acozryr (AT2020yxz) -- Spectroscopically confirmed afterglow of long GRB~201103B. ZTF20acozryr is described in detail as a case study in \S\ref{subsec:ZTF20acozryr}.

\item ZTF21aaarlbp (AT2021bl) -- Nova in the M33 galaxy.

\item ZTF21aabxjqr (SN 2021pb) -- Shock cooling of a Type IIb supernova at redshift $z=0.033$ (Fremling et al., in preparation).

\item ZTF21aaeyldq (AT2021any) -- Afterglow discovered serendipitously in ZTF data \citep{gcn29305} without any associated GRB. Spectroscopic observations allowed a redshift of $z=2.514$ to be measured and confirmed the nature of the transient \citep{gcn29307}. ZTF21aaeyldq was discovered as part of the high-cadence Partnership survey by a filter designed to find fast transients described in \citep{Ho2020, Perley2021arXiv}. The transient was missed during \texttt{ZTFReST} real-time operations because the time difference between the first and last detection in ZTF data was lower than our minimum baseline for light curve fitting of 0.5 days. Since the identification of ZTF21aaeyldq on 2021 January 16, we changed the threshold from 0.5 days to 0.125 days (3 hours). This allowed us to correctly recover ZTF21aaeyldq, which will be discussed in detail in future work (Ho et al., in preparation).

\item ZTF21aagwbjr (AT 2021buv) -- Confirmed afterglow of GRB~210204A \citep{gcn29405, gcn29408} at redshift $z = 0.876$ \citep{gcn29432}. ZTF photometry constrained the explosion time within 1.9 hours from the first detection. The transient was independently discovered also by the fast-transient filter described in \cite{Ho2020}. The light curve of ZTF21aagwbjr is shown in Figure\,\ref{fig:lightcurves bjr bav} and in Table~\ref{tab:ZTF21aagwbjr}. A dedicated work presenting multi-wavelength analysis of this source is planned (Kumar et al., in preparation).

\item ZTF21aahifke (AT2021clk) -- Fast transient found in ZTF data \citep{gcn29446} and rapidly confirmed with the GROWTH-India Telescope (GIT) follow-up observations (Figure~\ref{fig:lightcurve fke}). The nature of ZTF21aahifke, whose analysis is presented in \S\ref{subsec:ZTF21aahifke}, is still unknown.

\item ZTF21aapkbav (AT2021gca) -- Fast transient discovered on 2021 March 19 \citep{astronote_ZTF21aapkbav_disc} associated with the nearby galaxy GALEXMSC J142807.45+332950.0 at redshift $z=0.036$ \citep{KoEi2012}, as tabulated by the NASA/IPAC Extragalactic Database (NED). Under the assumption that ZTF21aapkbav and the galaxy are associated, the absolute magnitude of the transient's first (and brightest) detection was $M_r = -16.33 \pm 0.06$\,mag.
The fade rate of 0.3\,mag\,day$^{-1}$ that we measured from ZTF photometry (Figure\,\ref{fig:lightcurves bjr bav}, Table\,\ref{tab:ZTF21aapkbav}) in both $g$- and $r$- bands was confirmed by follow-up with the Lulin One-meter Telescope as part of the Kinder survey \citep{astronote_ZTF21aapkbav_kinder}.
The last ZTF upper limit before the first detection was measured on 2021 March 08 at 06:27 UT, which constrains the onset time to be within $< 9.1$ days from the first detection. Our team promptly triggered spectroscopy with Gemini North telescope equipped with the Gemini Multi-Object Spectrograph (GMOS) instrument.\footnote{Program ID GN-2021A-Q-102, PI Ahumada} The spectrum revealed a broad H$\alpha$ feature extending to velocity $v \sim 9,000$\,km\,s$^{-1}$ and possibly He I ($\lambda$=5876). The redshift was coarsely measured and consistent with the NED tabulated value. We concluded that ZTF21aapkbav was a fast-evolving supernova of Type IIb (see also Fremling et al., in preparation).

\end{itemize}

\section{Case studies}
\label{sec:case studies}

In this section, we present three case studies. Two are new, confirmed afterglows discovered by \texttt{ZTFReST}, namely ZTF20abtxwfx (AT2020sev) and ZTF20acozryr (AT2020yxz). The discovery process and GRB association will be described, along with multi-wavelength follow-up and data analysis. The third case study is a fast transient of unknown origin, ZTF21aahifke (AT2021clk).

\subsection{ZTF20abtxwfx}
\label{subsec:ZTF20abtxwfx}

The discovery of ZTF20abtxwfx (AT2020sev) occurred during science validation on 2020 August 28, which is about 10 days after the first detection in ZTF data, when \texttt{ZTFReST} was not yet being used in real time. During present operations, a transient like ZTF20abtxwfx would be deemed worthy of spectroscopic and photometric follow-up after the second night post-discovery (Figure \ref{fig:lightcurves}).

Multi-wavelength follow-up revealed a radio counterpart to ZTF20abtxwfx \citep{NaCh2020}, which suggests the transient to be a cosmological afterglow. Specifically, it is possible that ZTF20abtxwfx is the optical counterpart to GRB~200817A \citep{GCN_GRB200817A_GBM} or of a GRB that went undetected.
Here we briefly present optical, near-infrared, and radio follow-up of the transient, along with gamma--ray analysis carried out in a time frame where we could expect the onset of the event to be placed (including when GRB~200817A occurred).

\subsubsection{Optical and near-infrared}
\label{subsec:optical_ZTF20abtxwfx}

ZTF20abtxwfx was first detected on 2020 August 18 at 05:20 UT, heareafter labelled $T_{\text{det}}$. The ZTF forced-photometry optical light curve (Figure \ref{fig:lightcurves}; Table \ref{tab:ZTF20abtxwfx}) revealed a rapidly evolving transient that faded by $\sim1.3$\,mag in $r$-band in the first two days since $T_{\text{det}}$. The transient was last detected on 2020 August 23 at 04:51 UT, at $r = 20.98 \pm 0.24$ mag. Stringent upper limits constrained the transient onset time within $<1$ day from $T_{\text{det}}$. The color of the transient appeared to be red, with $g-r \sim 0.1$\,mag and $g-i \sim 0.3$\,mag one day after $T_{\text{det}}$. The Galactic extinction along the line of sight was low, with $E(B-V) = 0.015$\,mag \citep{Planck2014dust}.

We observed the source in the near-infrared with Palomar 200-inch (P200) equipped with Wide Field Infrared Camera (WIRC) on 2020 September 04 at 07:06 UT. The data were reduced using the automated pipeline described in \citep{De2020pasp}. No source was detected at the transient location, with 15 minutes of total exposure time, down to $J > 21.5$ ($5\sigma$).
Optical follow-up observations were obtained with LCO + 1-m Sinistro imager\footnote{Programs NOAO2020B-005, PI Coughlin; TOM2020A-008, PI Andreoni} and with P200 equipped with Wafer-Scale Imager for Prime (WaSP)\footnote{PI Kulkarni}.

A power-law fit of the ZTF $r$-band light curve, converted to flux and using the expression $f = f_0 (T - T_0)^{\alpha}$, returned an index $\alpha = -0.68 \pm 0.124$ and an estimated onset time $T_{\text{onset}}$ corresponding to 2020 August 17, 18:34:32 UT with a $1\sigma$ uncertainty of 4.08 hours \citep{GCN28355}. This value of $\alpha$ is within 2$\sigma$ of the mean of the $\alpha$-value distribution presented by \cite{del2016study}. Deep optical follow-up observations performed with P200+WaSP on 2020 September 17 at 03:12 UT, reduced with the pipeline described in \cite{De2020pasp}, significantly deviated from the power-law fit ($\sim 3\sigma$), which suggests that a jet break occurred.  This is evidence that, if ZTF20abtxwfx was the afterglow of a GRB, the GRB must have been on-axis, thus more easily detectable by space-based observatories than a GRB seen off-axis.

With a Bayesian analysis (Pang et al., in preparation) on the data with a light curve model of a GRB afterglow \citep{RyEe2020}, kilonova \citep{DiCo2020} and the combination of both, the evidence for these 3 hypotheses are estimated. The Bayes factors for GRB afterglow against the combination of GRB afterglow and kilonova is found to be $\sim10^{3.42\pm0.08}$. And the Bayes factor for GRB afterglow against kilonova is $\sim10^{3.58\pm0.08}$. Both suggest a strong preference for a GRB afterglow from the optical/NIR data alone.

\subsubsection{Radio}
\label{subsec:radio_ZTF20abtxwfx}

Radio follow-up observations of ZTF20abtxwfx were performed using the Very Large Array (VLA). We observed the field in X-band (central frequency 10\,GHz) twice, on 2020 August 31 and on 2020 September 15.
Data were calibrated using the automated pipeline available in the Common Astronomy Software Applications (CASA; \citealt{McWa2007}), with additional flagging performed manually, and imaged using the CLEAN algorithm \citep{Hog1974}.
On the first epoch, we found a point source spatially consistent with ZTF20abtxwfx. The flux density of the radio source was 50\,$\mu$Jy, with an image RMS of 4\,$\mu$Jy in 24 minutes of on-source time \citep{Ho2020ATel13986}. On the second epoch, acquired 15 days later, the flux density of the source decreased to $\sim$25\,$\mu$Jy.

A tentative detection with the upgraded Giant Metrewave Radio Telescope (uGMRT) was reported \citep{2020ATel14049....1N} at central frequency 1250\,MHz. \cite{2020ATel14049....1N} measured a flux density of $96 \pm 22 \mu$Jy on 2020 September 09 UT and $78 \pm 18 \mu$Jy on 2020 September 20 UT.

The presence of a fading radio counterpart supports the scenario in which ZTF20abtxwfx is the optical afterglow of a relativistic explosion. The measured decline of the radio light curve is consistent with what is commonly observed during GRB follow-up observations \citep[e.g.,][]{Chandra2012}.

\subsubsection{Gamma and X--rays}
\label{subsec:gamma_ZTF20abtxwfx}

We searched the GCN archives, the {\it Fermi}/GBM catalog, the {\it Fermi} GBM sub-threshold catalog, and the {\it Konus-Wind} triggered and waiting mode data for any possible counterpart in gamma--rays. We found one possible counterpart, GRB~200817A  \citep{GCN_GRB200817A_GBM}. The trigger time of GRB~200817A was 2020 August 17 at 09:25:20 UT, referred to as $T_0$ in this section. The GRB happened in the time interval between the last optical non-detection and the first detection of ZTF and within the $3\sigma$ time interval from $T_{\text{onset}}$. 

\begin{figure}
    \centering
    \includegraphics[width=\columnwidth]{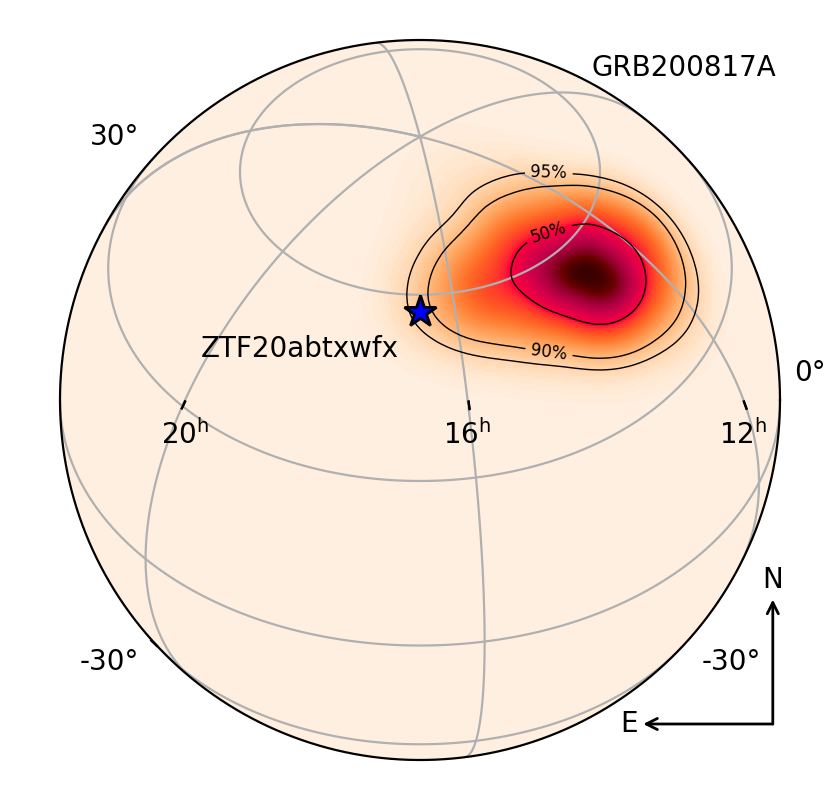}
    \caption{{\it Fermi}/GBM localization of GRB~200817A \citep{GCN_GRB200817A_GBM,goldstein2020evaluation}.  The optical transient ZTF20abtxwfx, marked with a blue star in the map, is included in the 93th percentile of the localization probability. The figure was created using the \texttt{ligo.skymap} Python package.}
    \label{fig:ZTF20abtxwfx_loc}
\end{figure}

\begin{figure}
    \centering
    \includegraphics[width=\columnwidth]{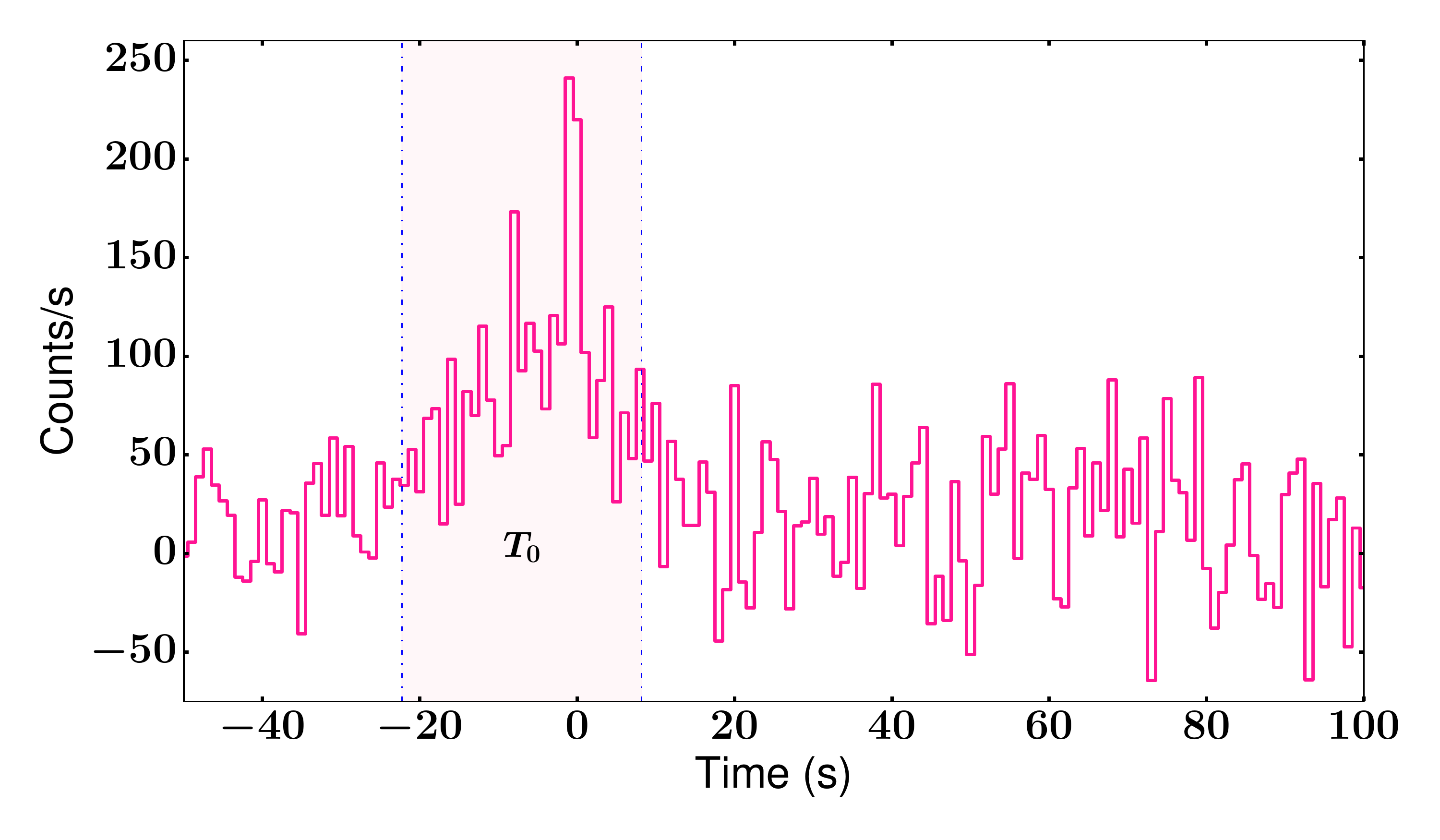}
    \includegraphics[height=7cm]{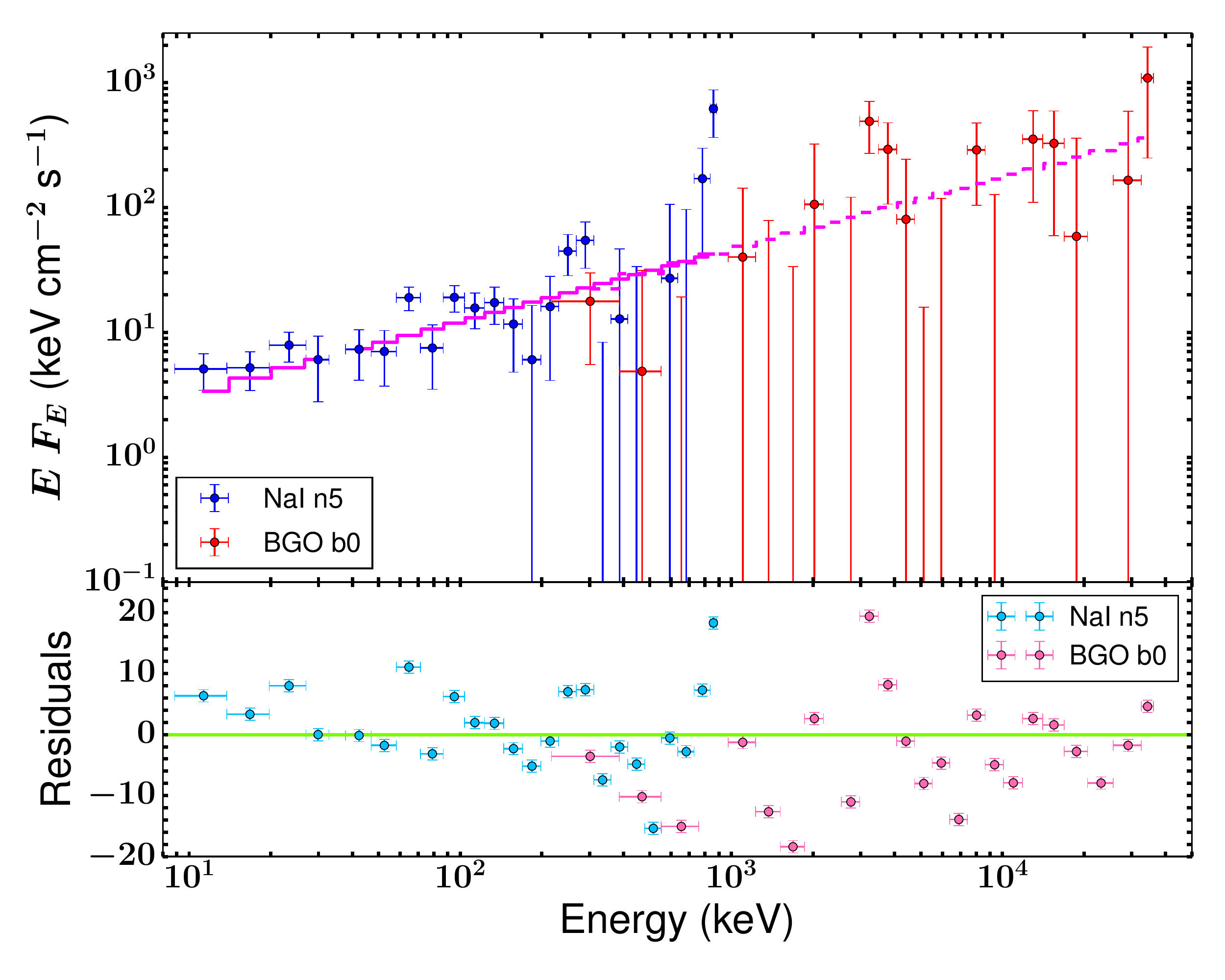}
    \caption{{\it Fermi}/GBM background subtracted lightcurve for GRB 200817A using data from NaI detector n5 in the energy range 8 - 900 keV. The top panel shows the emission episodes used for the time-integrated spectral analysis covering $-$22.27\,s to 8.19\,s. The bottom panel shows the broadband spectroscopy for the prompt emission episode of GRB 200817A fitted with a simple power-law. The blue and red data points are for NaI n5 and BGO b0 detectors respectively.}
    \label{fig:GRB200817A_spectra}
\end{figure}

ZTF20abtxwfx was located in the 93$^{\text{rd}}$ percentile of the {\it Fermi} localization of GRB~200817A \citep[Figure\,\ref{fig:ZTF20abtxwfx_loc};][]{GCN_GRB200817A_GBM, goldstein2020evaluation}. 
We confirmed that the position of ZTF20abtxwfx was within the GBM field of view at the time of GRB~200817A.
The initial automated classification flag identified this burst as a short GRB; however, this was due to an unusual slow rise followed by a sharp spike which resulted in an incorrect source interval selection by the automated processing. The final duration measure of $T_{90}=30.46$\,s results in a long GRB classification for GRB~200817A (Figure\,\ref{fig:GRB200817A_spectra}).

We performed a spectral analysis of GRB~200817A using {\it Fermi}/GBM data. 
We chose data from GBM detectors so that the boresight angle was $< 50$\,deg. For GRB~200817A, the selected detectors were n5 ($21$\,deg) and b0 ($21$\,deg). The GRB spectra were analyzed in a time interval of 30.46\,s, equal to $T_{90}$,
from $T_0-22.27$\,s to $T_0 + 8.19$\,s. We chose pre- and post-time intervals of 100\,s for the background measurement. We have fit the data with several models including simple power law, cutoff power-law, Band function, and GRBCOMP model. 
A simple power-law model was found to be the best fit model, with power-law index $1.45(\pm 0.04)$ and with E$_{\text{norm}}$ constrained to 100\,keV (Figure\,\ref{fig:GRB200817A_spectra}). The model flux in the 10--1000\,keV range was found to be $1.19\times10^{-7}$ ergs\,cm$^{-2}$\,s$^{-1}$.

Furthermore, we carried out a thorough search for gamma--ray counterparts to ZTF20abtxwfx in data acquired with {\it AstroSat} Cadmium Zinc Telluride Imager (CZTI).
ZTF20abtxwfx was not Earth-occulted at the time of {\it Fermi} GRB~200817A as seen from {\it AstroSat}. However, when we calculated fluxes using the parameters inferred from {\it Fermi}, we concluded that it was too faint for {\it AstroSat} to detect it, since we calculated an {\it AstroSat} dectection limit of $\sim 3\times10^{-7}$\,erg\,cm$^{-2}$\,s$^{-1}$ for 4s binning in the direction of the transient. Therefore we cannot rule out an association between ZTF20abtxwfx and GRB~200817A, despite the {\it AstroSat} non-detection. 

Finally, we conducted a search for
new bursts in a time window of $T_{\text{onset}} \pm 12$ hours, which approximately corresponds to the 3$\sigma$ time interval from the expected onset of ZTF20abtxwfx inferred from the optical light curve (see \S\ref{subsec:optical_ZTF20abtxwfx}). The search did not yield any significant detection in CTZI data \citep{GCN28355}. The CTZI burst closest to $T_{\text{onset}}$ was GRB~200817B, which was detected on 2020 August 17 06:03:44 UT \citep{GCN28354}, outside the $3\sigma$ time interval from $T_{\text{onset}}$. We conclude that GRB~200817A is the most likely high-energy counterpart to ZTF20abtxwfx.

\subsection{ZTF20acozryr: serendipitous discovery of a long GRB afterglow in real time}
\label{subsec:ZTF20acozryr}

 The optical fast transient candidate ZTF20acozryr (AT2020yxz) was flagged by the \texttt{ZTFReST} pipeline on 2020 November 05 \citep{CoAn2020b}. ZTF20acozryr was first detected by ZTF on 2020 November 03 at 09:44 UT and faded by $\sim$\,0.7\,mag in $g$-band in $\sim$\,0.9 days.
A $g$-band upper limit constrained the onset time to be within $\lesssim 1$\,day of the first detection.
The photometry for this object is presented in Figure\,\ref{fig:lightcurves} and Table~\ref{tab:ZTF20acozryr}. The initial color of the transient appeared to be relatively red, with $g-r \sim 0.3$\,mag around the time of the first observation. The Galactic extinction on the line of sight, $E(B-V)=0.10$\,mag, was too low  to be responsible for the red color.

 The rapid fade rate was flagged in near real time and LCO follow-up observations were promptly triggered\footnote{Programs NOAO2020B-005, PI Coughlin; TOM2020A-008, PI Andreoni}. These observations were crucial for continued light curve sampling for this object, given its faintness upon detection ($\sim$\,19.5 in $g$-band, $\sim$\,19.2 in $r$-band) and the measured fade rate above 0.5\,mag per day. Early discovery allowed us and the community to promptly trigger follow-up observations: spectroscopy from the Very Large Telescope (VLT) $\sim 2.3$\,days later led to a measured redshift $z = 1.105$, spectroscopically confirming ZTF20acozryr to be an afterglow \citep{XuVi2020}.

At the same time, a Swift ToO observation (ID 21039) was approved, resulting in a confirmation of an X--ray counterpart \citep{EvKu2020}. Similarly, the IPN network confirmed consistency of the transient's location with the localization inferred for the long-duration, bright GRB~201103B \citep{SvGo2020} first reported by AGILE \citep{UrVi2020}.

After its prompt public announcement \citep{CoAn2020b}, ZTF20acozryr was imaged with several telescopes. Photometric follow-up was reported by \cite{gcn28846, gcn28854, gcn28862, gcn28875, gcn28876, gcn28880, gcn28883, gcn28886, gcn28925, gcn28926, gcn29181}; see also Figure\,\ref{fig:lightcurves}. 

On 2021 January 11, about 10 weeks after GRB~201103B, we observed ZTF20acozryr with the Low Resolution Imaging Spectrometer \citep[LRIS;][]{Oke1995} at W. M. Keck Observatory. Data were reduced using \texttt{lpipe} \citep{Perley2019lpipe}, a fully automatic
data reduction pipeline for imaging and spectroscopy. The transient was not detected down to $r \sim g > 25.5$\,mag (see the inset plot in Figure\,\ref{fig:lightcurves}).

We again performed a power-law fit in the form $f = f_0 (T - T_0)^{\alpha}$, where $T_0$ was fixed to be the discovery time of GRB~201103B. For the fit, we used data acquired by our team with ZTF and during the follow-up of ZTF20acozryr with GIT, LCO, and Liverpool Telescope (LT). The full dataset can be described by a power-law with index $\alpha = -1.14 \pm 0.13$. However, the low {\it p}-value of the fit ($p < 10^{-5}$) suggests that the data deviate significantly from the power-law curve. A better fit ($p \sim 0.4$) is obtained by ignoring the GROWTH-India Telescope (GIT) data point taken $\sim 6$ days after the GRB occurred. In this case, the data can be fit by a power-law with index $\alpha = -0.96 \pm 0.06$. We suggest that a jet-break occurred $\sim 4-5$ days after the GRB. This scenario is consistent with reported photometry \citep{gcn28875, gcn28880, gcn28926, gcn29181} and with LRIS late-time data.

We perform a Bayesian analysis similar to what was done for ZTF20abtxwfx. The Bayes factors for GRB afterglow against the combination of GRB afterglow and kilonova is found to be $\sim10^{0.21\pm0.09}$. And the Bayes factor for GRB afterglow against kilonova is $\sim10^{1532.9\pm0.15}$. The extremely low evidence for the data originating from a kilonova is due to the high redshift of the detection. Again, the data suggest a GRB afterglow origin for the optical transient.

\subsection{ZTF21aahifke: a fast transient of unknown origin}
\label{subsec:ZTF21aahifke}

\begin{figure*}
    \centering
    \includegraphics[width=0.8\textwidth]{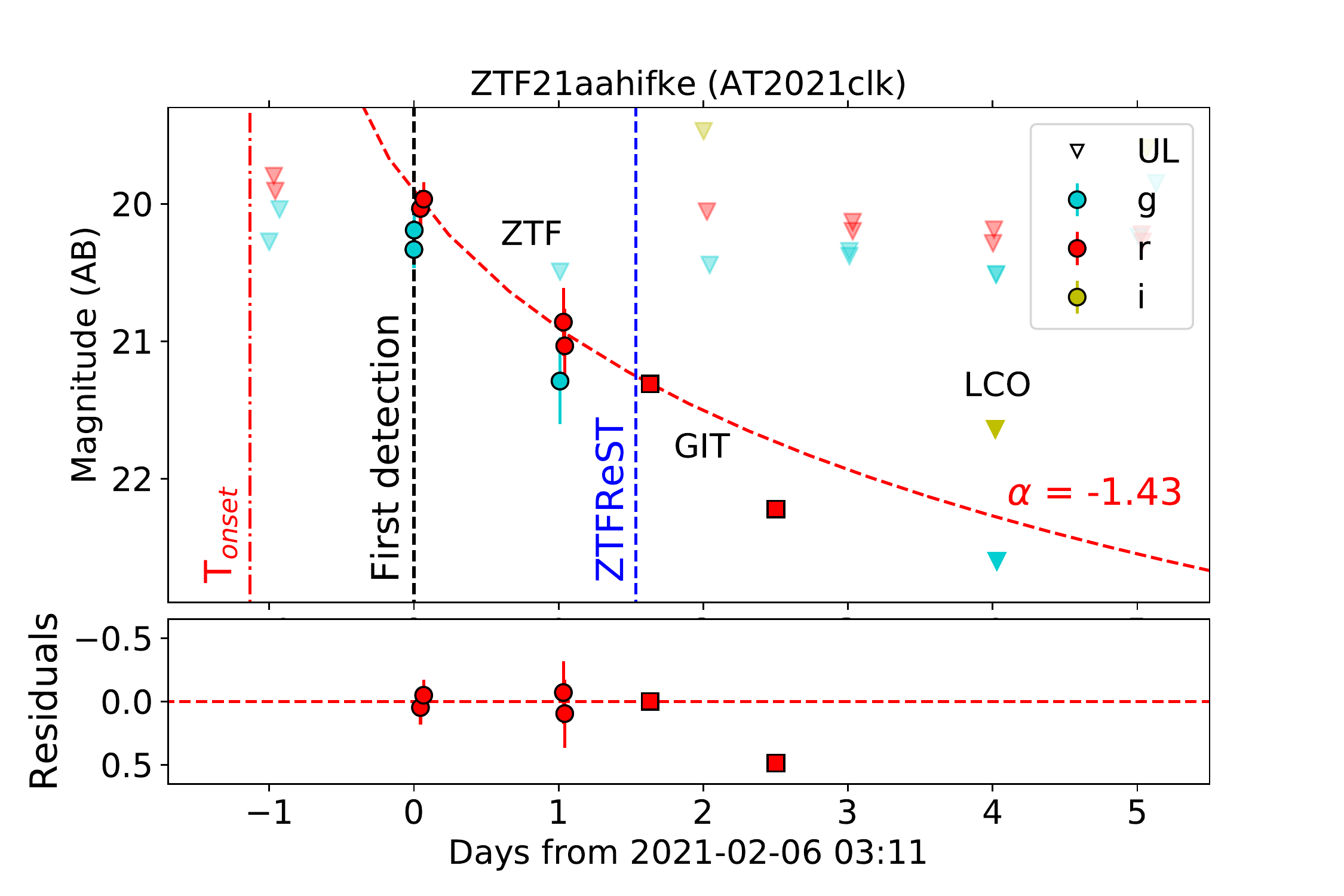}
    \caption{Photometry light curve of the fast transient ZTF21aahifke, which does not have any known associated gamma-ray counterpart. The power-law fit to the ZTF detections and the first GIT data returns results with very large uncertainties. The resulting onset time ($T_{\text{onset}}$) is placed on 2021 February 4 23:57 (before the last ZTF non-detection preceding the discovery of ZTF21aahifke), with a standard deviation of 1.57 days, and the power-law index is $\alpha = -1.43 \pm 1.27$.}
    \label{fig:lightcurve fke}
\end{figure*}

 The optical fast transient candidate ZTF21aahifke (AT2021clk) was identified as a rapidly declining source on 2021 February 07, with a fade rate of $0.96$\,mag\,day$^{-1}$ \citep{gcn29446}. The discovery of ZTF21aahifke as a fast-fading source was 
made possible by forced photometry on the first night after its $>5\sigma$ detection. Photometric follow-up with GIT confirmed the transient detection as well as the rapid decline of its brightness (Figure\,\ref{fig:lightcurve fke}; Table\,\ref{tab:ZTF21aahifke}). LCO photometry (TOM2020A-008, PI Andreoni) in $g$- and $i$-band 4 days from the first detection placed upper limits that ruled out a significant re-brightening, or extreme optical colors, of the source. At the time of first detection, the color of ZTF21aahifke appears to be red, with $g-r \sim 0.3$\,mag. The Galactic extinction on the line of sight is $E(B-V)=0.12$\,mag \citep{Planck2014dust}, too low to explain the red colors.

 We used forced photometry to search for previous activity in 1,109 ZTF epochs taken before the 2020 February 06, without finding any significant detection.
Previous activity was also searched for and not found in the Pan-STARRS
\citep{ChMa2016} Data Release 2 catalog and in ATLAS images, explored via the
public forced photometry server \citep{ToDe2018,SmSm2020}.
Deep Pan-STARRS images did not reveal any underlying source at the transient location. The closest source is the SDSS J025427.89+363151.5 galaxy, $\sim 6.6''$ away from ZTF21aahifke, with Sloan Digital Sky Survey (SDSS) photometric redshift of $z = 0.433 \pm 0.1735$; at this redshift, such an offset is unlikely for afterglows.
We found no {\it Fermi} or {\it Swift} gamma-ray burst alert, issued in the 22.3 hours between the last non-detection and the first detection of ZTF21aahifke, with compatible localization (95\%). GIT follow-up photometry shows a steepening of the light curve (Figure\,\ref{fig:lightcurves bjr bav}) which could suggest the presence of a jet break, under the assumption that ZTF21aahifke is another un-triggered GRB afterglow.

On 2021 February 20, one epoch of radio data was acquired with VLA (PI Perley). The data were reduced in the same way as for ZTF20abtxwfx. No radio counterpart to ZTF21aahifke was found, $\sim 15$ days from the transient onset, with an RMS of 7$\mu$Jy. 

We also perform a Bayesian analysis similar to what was done for ZTF20abtxwfx. Leaving the distance as a free parameter, the Bayes factors for a kilonova against a GRB afterglow is found to be $\sim10^{4.37\pm0.08}$. And the Bayes factor for kilonova against GRB afterglow plus kilonova is $\sim10^{0.04\pm0.06}$. The moderate favoring for the kilonova hypothesis is due to the non-detection at 5\,days, which is mildly constraining to the afterglow model best fit to the early points, and does not have a potential jet break encoded, while the kilonova model has dropped well below the limit by that time.
Although ZTF21aahifke shares similarities with other afterglows observed with ZTF, we cannot confidently exclude a kilonova or a Galactic origin, such as from a cataclysmic variable and other Galactic fast transients identified in the survey \citep{AnKo2020}.

\section{Kilonova rate}
\label{sec:kilonova_rate}
Real-time searches for extragalactic fast transients make it possible to determine the rates of kilonovae, improving the results obtained by \cite{AnKo2020} and approaching those of \cite{KaNa2017}, and therefore the rates of binary neutron star (and neutron star--black hole) mergers. 

In this work, we combined results from archival and real-time searches. During 7 months of archival searches with a cut at Galactic latitude $b_{\text{Gal}} > 10$\,deg, the number of fields explored is consistent with \cite{AnKo2020} within 1\%. During 5 months of real-time operation, no cut in Galactic latitude was applied, thus 23\% more fields were explored than in \cite{AnKo2020}, albeit with higher Galactic extinction and in typically more crowded fields. Thus, the following results can be considered conservative.

\cite{AnKo2020} constrained the GW170817-like kilonova rate to be $R < 1775$\,Gpc$^{-3}$yr$^{-1}$ (95\% confidence) with 23 months of ZTF survey data. In this paper, we have analyzed more than 12 additional months of ZTF survey data with methods more conservative than those used by \cite{AnKo2020}; we found no confirmed kilonova. 

As in \cite{AnKo2020}, we used \texttt{simsurvey} \citep{FeNo2019} to inject kilonova light curves in ZTF survey data and infer the kilonova rate by measuring the number of synthetic kilonovae recovered by the software.
The new analysis ranged data spanning 2020 February 22 to 2021 March 03.
To be ``detected,'' we required that recovered lightcurves have a fade rate larger than 0.3\,mag in each band; in addition, we required at least two detections with $>3\sigma$ significance, at least one of which must have $>5 \sigma$ significance, and $>3$ hours of time separation. 
The new set of data alone, which was characterized by the higher cadence of ZTF Phase II, gives a rate constraint of 
$R < 1904$\,Gpc$^{-3}$\,yr$^{-1}$; combined with the previous analysis, this provides a rate constraint $R < 900$\,Gpc$^{-3}$\,yr$^{-1}$ for kilonovae similar to GW170817. This measurement improves our previous limits by 49\%. To demonstrate the sensitivity of our results to our assumptions, a limit requiring two or more detections with $5\sigma$ significance yields $R_{\text{KN}} < 1017$\,Gpc$^{-3}$yr$^{-1}$, less constraining by a small amount.

\begin{figure*}
    \centering
    \includegraphics[width=\textwidth]{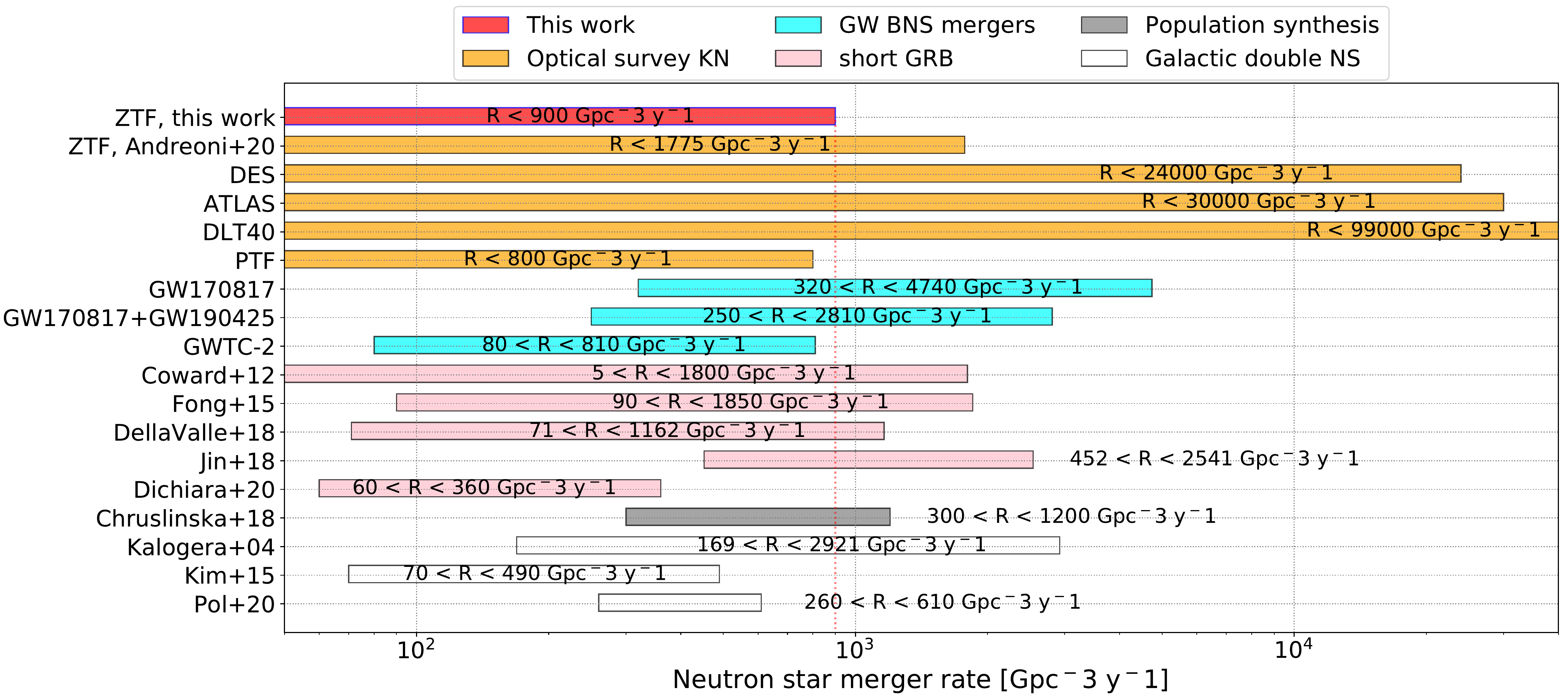}
    \caption{Upper limits on kilonova rates from optical surveys \citep[this work and][]{AnKo2020, Doctor2017, KaNa2017, Smartt2017, Yang2017}  compared against binary neutron star merger rates obtained via GW \citep{LIGO_GWTC2_pop_2020arXiv, PhysRevLett.119.161101, AbEA2019}, short GRB \citep{Coward2012, Fong15, DellaValle2018, Jin2018, Dichiara2020}, and Galactic double neutron star observations \citep{Kalogera2004erratum, Kim2015, Pol2020}, along with population synthesis results \citep{Chruslinska2018}. 
    }
    \label{fig:rates}
\end{figure*}

An upper limit of $R_{\text{KN}} < 900$\,Gpc$^{-3}$yr$^{-1}$ for GW170817-like kilonovae is consistent (Figure \ref{fig:rates}) with the most recent binary neutron star merger rate, inferred from GW observations, of
$R_{\text{BNS}} = 320^{+490}_{-240}$\,Gpc$^{-3}$yr$^{-1}$ \citep{LIGO_GWTC2_pop_2020arXiv}; our limits may be optimistic as kilonovae may exist that are fainter than GW170817 \citep[for a recent analysis of the kilonova luminosity function based on O3 follow-up observations, see][]{KaAn2020}.

\begin{figure}[ht]
    \centering
    \includegraphics[width=\columnwidth]{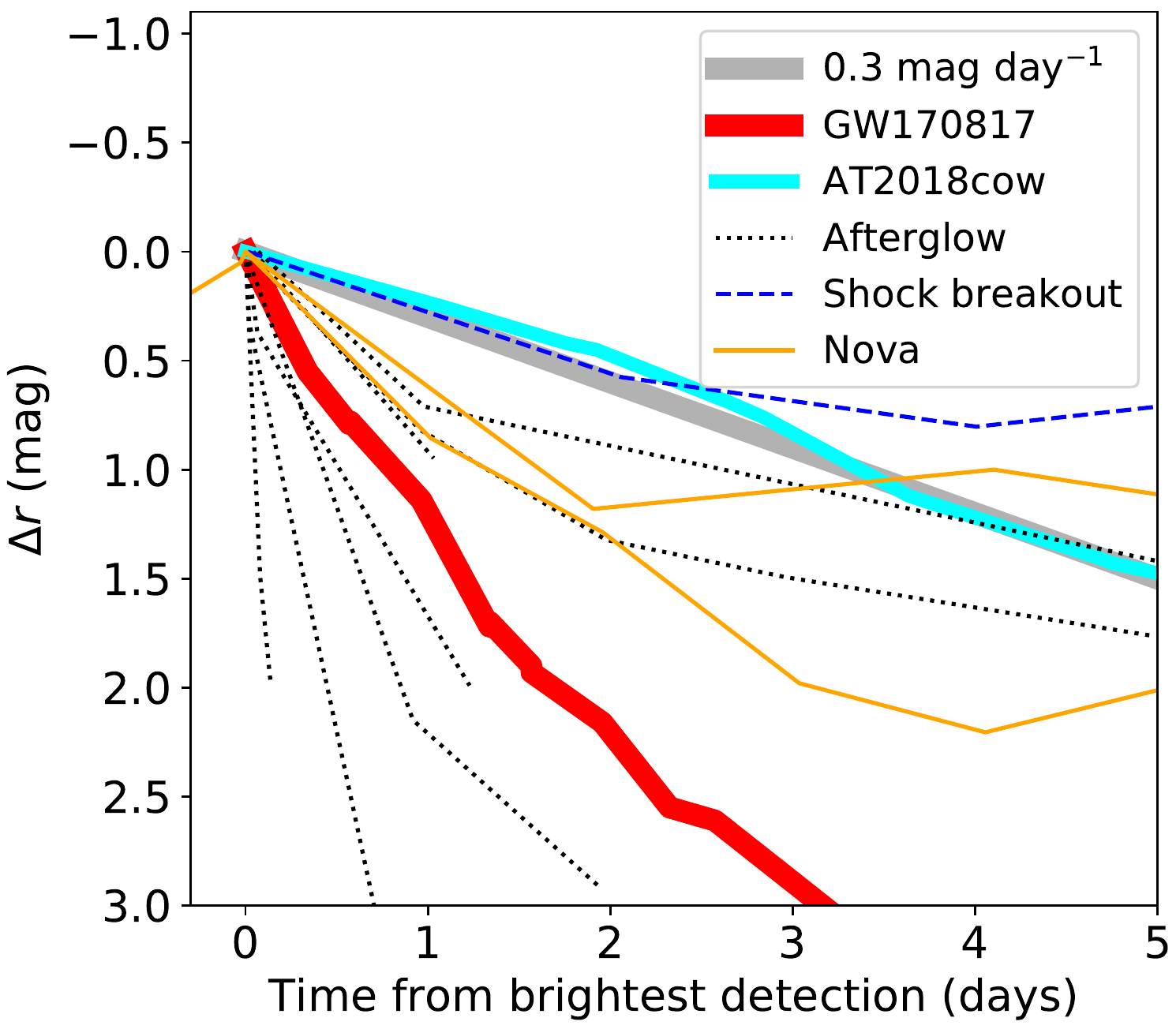}
    \caption{Evolution in $r$-band of extragalactic fast transients found by \texttt{ZTFReST}, compared with the GW170817 kilonova \citep[data from][]{ArHo2017, KaNa2017, SmCh2017} and the FBOT AT2018cow \citep[data from][]{PeMa2019cow}. }
    \label{fig:lc_compare}
\end{figure}

\section{Conclusions}
\label{sec:conclusion}

In this paper, we have presented an overview of \texttt{ZTFReST}, an open-source infrastructure built on top of the ZTF alert stream to search for kilonovae and other fast transients. We described how this infrastructure has already yielded a number of candidates, including at least seven confirmed afterglows. Four of these were found during 265 days of science validation on archival searches (ZTF20aajnksq, ZTF20abbiixp, ZTF20abwysqy, ZTF20abtxwfx),
two were discovered during real-time operations (ZTF20acozryr and ZTF21aagwbjr), and one was recovered after a parameter optimization of the pipeline (ZTF21aaeyldq).
The \texttt{ZTFReST} early identification of three of these afterglows, specifically ZTF20abtxwfx, ZTF20acozryr, and ZTF21aagwbjr, made it possible to carry out multi-wavelength follow-up observations.

Kilonovae are rapid transients of primary importance for this project. The non-detection of viable kilonovae in the ZTF dataset allowed us to constrain the GW170817-like kilonova rate to $R < 900$\,Gpc$^{-3}$\,yr$^{-1}$. We found that cosmological afterglows (with or without a gamma--ray counterpart) are the dominant ``contaminants" for kilonova searches at high Galactic latitude, after cataclysmic variables and flare stars are rejected via accurate vetting.
Figure\,\ref{fig:lc_compare} presents a comparison between the re-scaled early light curves of the extragalactic fast transients found by \texttt{ZTFReST}, the GW170817 kilonova, and the fast blue optical transient (FBOT) AT2018cow.
This shows the potential of \texttt{ZTFReST} for serendipitous afterglow discovery, i.e. independent of gamma-ray triggers. 
More of such discoveries could eventually shed some light on the puzzling paucity of ``dirty fireballs'' and ``orphan'' GRB afterglows \citep{Dermer2000, Huang2002, Rhoads2003} in optical surveys.  Future work will present more details on the afterglows discovered by ZTF, with a discussion on how the event rate compares to expectations for long GRBs and the implications for hypothesized populations of dirty fireballs. 

This effort will also inform us about what strategies, both in terms of survey and alert characterization, will improve kilonova searches going forward. For example, both the cadence and filter choices of ZTF can change the stream of transient candidates that passes our thresholds \citep[see also][]{AlCo2020arXiv}.

Future synoptic surveys such as Vera C. Rubin Observatory's LSST will rely on alert streams very similar to ZTF. LSST is expected to produce $\sim 10$M alerts per night, which presents us with a data mining challenge, but also puts a strain on follow-up telescope resources.
 While LSST will perform forced photometry on all transients, image-stacking services are currently not planned for LSST, although they could be key to unveiling a population of tens of kilonovae, especially in fields observed with nightly cadence \citep{Andreoni2019LSST}. Based on results from ZTF, we can expect the application of \texttt{ZTFReST} to the LSST alert stream to yield a manageable number of extragalactic fast transients with a low number of ``false positives" outside of the Galactic plane, especially if cross-match with nearby galaxies is required. A dedicated performance analysis with LSST test alerts, based on DECam optical observations, is planned. 
 The transients found by LSST and selected with \texttt{ZTFReST} can then be prioritized for rapid characterization with large telescopes. This approach may represent the scientifically crucial divide between candidate {\it detection} and transient {\it discovery} in the LSST era.
\\

\acknowledgments
\section*{Acknowledgments}
M.~W.~C acknowledges support from the National Science Foundation with grant number PHY-2010970.
M. B. acknowledges support from the Swedish Research Council (Reg. no. 2020-03330).
A.S.C, E. C. K., A.G and J.S,  acknowledge support from the G.R.E.A.T. research environment funded by {\em Vetenskapsr\aa det}, the Swedish Research Council, under project number 2016-06012, and support from The Wenner-Gren Foundations.

\noindent Bayes factor computations between kilonova and GRB afterglow models have used computational resources provided through SuperMUC\_NG (LRZ) under project number pn29ba and Hawk (HLRS) under project number 44189.

\noindent This work was supported by the GROWTH (Global Relay of Observatories Watching Transients Happen) project funded by the National Science Foundation under PIRE Grant No 1545949. GROWTH is a collaborative project among California Institute of Technology (USA), University of Maryland College Park (USA), University of Wisconsin Milwaukee (USA), Texas Tech University (USA), San Diego State University (USA), University of Washington (USA), Los Alamos National Laboratory (USA), Tokyo Institute of Technology (Japan), National Central University (Taiwan), Indian Institute of Astrophysics (India), Indian Institute of Technology Bombay (India), Weizmann Institute of Science (Israel), The Oskar Klein Centre at Stockholm University (Sweden), Humboldt University (Germany), Liverpool John Moores University (UK) and University of Sydney (Australia).

\noindent Based on observations obtained with the Samuel Oschin Telescope 48-inch and the 60-inch Telescope at the Palomar Observatory as part of the Zwicky Transient Facility project. ZTF is supported by the National Science Foundation under Grant No. AST-2034437 and a collaboration including Caltech, IPAC, the Weizmann Institute for Science, the Oskar Klein Center at Stockholm University, the University of Maryland, Deutsches Elektronen-Synchrotron and Humboldt University, the TANGO Consortium of Taiwan, the University of Wisconsin at Milwaukee, Trinity College Dublin, Lawrence Livermore National Laboratories, and IN2P3, France. Operations are conducted by COO, IPAC, and UW. The ZTF forced-photometry service was funded under the Heising-Simons Foundation grant \#12540303 (PI: Graham).

\noindent This work has made use of data from the Asteroid Terrestrial-impact Last Alert System (ATLAS) project. The Asteroid Terrestrial-impact Last Alert System (ATLAS) project is primarily funded to search for near earth asteroids through NASA grants NN12AR55G, 80NSSC18K0284, and 80NSSC18K1575; byproducts of the NEO search include images and catalogs from the survey area. This work was partially funded by Kepler/K2 grant J1944/80NSSC19K0112 and HST GO-15889, and STFC grants ST/T000198/1 and ST/S006109/1. The ATLAS science products have been made possible through the contributions of the University of Hawaii Institute for Astronomy, the Queen’s University Belfast, the Space Telescope Science Institute, the South African Astronomical Observatory, and The Millennium Institute of Astrophysics (MAS), Chile.

\bibliographystyle{aasjournal}
\bibliography{references}

\clearpage
\begin{appendix}

\section{Photometry Tables}

\begin{longtable}{cccccc}
    \hline
    \hline
    Instrument & Time (UT) & Mag (AB) & Magerr (AB) & Lim mag (AB) & Filter \\
\hline
ZTF & 2020-08-16 05:11 & 99.00 & 99.00 & 20.5 & g \\
ZTF & 2020-08-16 06:12 & 99.00 & 99.00 & 20.4 & r \\
ZTF & 2020-08-17 05:15 & 99.00 & 99.00 & 20.6 & g \\
ZTF & 2020-08-17 06:44 & 99.00 & 99.00 & 20.3 & r \\
ZTF & 2020-08-18 05:19 & 19.14 & 0.05 & 20.4 & r \\
ZTF & 2020-08-19 04:19 & 19.90 & 0.12 & 20.0 & r \\
ZTF & 2020-08-19 05:23 & 20.14 & 0.11 & 20.4 & g \\
ZTF & 2020-08-19 05:53 & 19.79 & 0.17 & 19.6 & i \\
ZTF & 2020-08-20 04:46 & 20.49 & 0.19 & 20.2 & r \\
ZTF & 2020-08-20 05:46 & 20.66 & 0.19 & 20.3 & g \\
ZTF & 2020-08-21 04:24 & 20.68 & 0.21 & 20.2 & r \\
ZTF & 2020-08-21 06:53 & 21.16 & 0.34 & 20.1 & g \\
ZTF & 2020-08-22 03:42 & 99.00 & 99.00 & 17.3 & g \\
ZTF & 2020-08-23 04:12 & 99.00 & 99.00 & 20.3 & g \\
ZTF & 2020-08-23 04:50 & 20.79 & 0.25 & 20.1 & r \\
ZTF & 2020-08-24 04:25 & 99.00 & 99.00 & 19.8 & r \\
ZTF & 2020-08-24 06:38 & 99.00 & 99.00 & 19.7 & i \\
ZTF & 2020-08-25 04:15 & 99.00 & 99.00 & 20.3 & r \\
ZTF & 2020-08-25 04:45 & 99.00 & 99.00 & 20.3 & g \\
ZTF & 2020-08-26 03:24 & 99.00 & 99.00 & 19.9 & r \\
ZTF & 2020-08-26 03:49 & 99.00 & 99.00 & 20.0 & r \\
ZTF & 2020-08-26 04:16 & 99.00 & 99.00 & 20.0 & g \\
ZTF & 2020-08-27 04:13 & 99.00 & 99.00 & 19.8 & r \\
ZTF & 2020-08-27 06:09 & 99.00 & 99.00 & 19.5 & g \\
ZTF & 2020-08-28 04:35 & 99.00 & 99.00 & 19.6 & i \\
ZTF & 2020-08-28 05:12 & 99.00 & 99.00 & 19.4 & r \\
ZTF & 2020-08-28 05:41 & 99.00 & 99.00 & 19.3 & g \\
ZTF & 2020-08-29 04:16 & 99.00 & 99.00 & 20.0 & r \\
ZTF & 2020-08-29 05:41 & 99.00 & 99.00 & 19.8 & g \\
ZTF & 2020-09-01 04:22 & 99.00 & 99.00 & 19.5 & g \\
ZTF & 2020-09-01 04:23 & 99.00 & 99.00 & 19.5 & g \\
ZTF & 2020-09-01 05:20 & 99.00 & 99.00 & 19.5 & r \\
ZTF & 2020-09-01 05:48 & 99.00 & 99.00 & 19.2 & i \\
ZTF & 2020-09-02 06:47 & 99.00 & 99.00 & 19.2 & r \\
ZTF & 2020-09-02 06:49 & 99.00 & 99.00 & 19.3 & r \\
Sinistro & 2020-09-03 02:43 & 99.00 & 99.00 & 21.7 & g \\
Sinistro & 2020-09-03 02:51 & 99.00 & 99.00 & 21.9 & r \\
Sinistro & 2020-09-03 03:00 & 99.00 & 99.00 & 21.6 & i \\
ZTF & 2020-09-03 03:46 & 99.00 & 99.00 & 19.9 & r \\
ZTF & 2020-09-03 03:47 & 99.00 & 99.00 & 19.9 & r \\
ZTF & 2020-09-03 04:15 & 99.00 & 99.00 & 19.6 & g \\
ZTF & 2020-09-03 04:20 & 99.00 & 99.00 & 19.6 & g \\
ZTF & 2020-09-04 04:26 & 99.00 & 99.00 & 19.9 & g \\
ZTF & 2020-09-04 04:27 & 99.00 & 99.00 & 19.9 & g \\
ZTF & 2020-09-04 04:50 & 99.00 & 99.00 & 20.0 & r \\
ZTF & 2020-09-04 05:20 & 99.00 & 99.00 & 20.0 & i \\
ZTF & 2020-09-04 06:24 & 99.00 & 99.00 & 20.0 & i \\
ZTF & 2020-09-04 06:45 & 99.00 & 99.00 & 19.6 & r \\
ZTF & 2020-09-04 06:47 & 99.00 & 99.00 & 19.7 & r \\
WIRC & 2020-09-04 07:06 & 99.00 & 99.00 & 21.5 & J \\
ZTF & 2020-09-05 04:51 & 99.00 & 99.00 & 19.8 & g \\
ZTF & 2020-09-05 05:13 & 99.00 & 99.00 & 19.9 & i \\
ZTF & 2020-09-05 05:16 & 99.00 & 99.00 & 19.6 & i \\
ZTF & 2020-09-05 05:45 & 99.00 & 99.00 & 19.8 & i \\
ZTF & 2020-09-05 06:46 & 99.00 & 99.00 & 19.6 & r \\
ZTF & 2020-09-06 03:52 & 99.00 & 99.00 & 20.5 & r \\
ZTF & 2020-09-06 03:55 & 99.00 & 99.00 & 20.4 & r \\
ZTF & 2020-09-06 06:23 & 99.00 & 99.00 & 19.9 & i \\
ZTF & 2020-09-06 06:37 & 99.00 & 99.00 & 19.6 & i \\
ZTF & 2020-09-07 03:43 & 99.00 & 99.00 & 20.6 & g \\
ZTF & 2020-09-07 03:44 & 99.00 & 99.00 & 20.6 & g \\
ZTF & 2020-09-07 03:49 & 99.00 & 99.00 & 20.6 & g \\
ZTF & 2020-09-07 04:44 & 99.00 & 99.00 & 20.4 & r \\
ZTF & 2020-09-07 04:45 & 99.00 & 99.00 & 20.4 & r \\
ZTF & 2020-09-07 05:23 & 99.00 & 99.00 & 20.1 & r \\
ZTF & 2020-09-07 05:43 & 99.00 & 99.00 & 19.9 & i \\
ZTF & 2020-09-07 06:52 & 99.00 & 99.00 & 19.5 & i \\
ZTF & 2020-09-14 03:40 & 99.00 & 99.00 & 19.1 & i \\
ZTF & 2020-09-14 04:17 & 99.00 & 99.00 & 19.7 & g \\
ZTF & 2020-09-14 04:50 & 99.00 & 99.00 & 19.7 & r \\
ZTF & 2020-09-14 05:57 & 99.00 & 99.00 & 19.4 & i \\
ZTF & 2020-09-14 06:11 & 99.00 & 99.00 & 19.4 & i \\
ZTF & 2020-09-15 03:27 & 99.00 & 99.00 & 19.7 & r \\
ZTF & 2020-09-15 03:53 & 99.00 & 99.00 & 19.8 & g \\
ZTF & 2020-09-15 04:18 & 99.00 & 99.00 & 19.7 & g \\
ZTF & 2020-09-15 04:50 & 99.00 & 99.00 & 19.6 & i \\
ZTF & 2020-09-15 05:52 & 99.00 & 99.00 & 19.6 & i \\
ZTF & 2020-09-15 06:15 & 99.00 & 99.00 & 19.3 & r \\
ZTF & 2020-09-17 02:56 & 99.00 & 99.00 & 20.1 & i \\
ZTF & 2020-09-17 03:11 & 99.00 & 99.00 & 20.1 & i \\
WaSP & 2020-09-17 03:12 & 23.86 & 0.16 & 24.0 & i \\
WaSP & 2020-09-17 03:32 & 23.95 & 0.16 & 24.0 & r \\
ZTF & 2020-09-17 03:42 & 99.00 & 99.00 & 20.2 & r \\
ZTF & 2020-09-17 04:07 & 99.00 & 99.00 & 20.0 & r \\
ZTF & 2020-09-17 04:43 & 99.00 & 99.00 & 20.0 & g \\
ZTF & 2020-09-17 05:09 & 99.00 & 99.00 & 19.9 & g \\
ZTF & 2020-09-18 03:26 & 99.00 & 99.00 & 20.3 & r \\
ZTF & 2020-09-18 03:44 & 99.00 & 99.00 & 20.3 & g \\
ZTF & 2020-09-18 04:09 & 99.00 & 99.00 & 19.6 & i \\
ZTF & 2020-09-18 04:11 & 99.00 & 99.00 & 20.1 & i \\
ZTF & 2020-09-18 04:46 & 99.00 & 99.00 & 19.7 & i \\
ZTF & 2020-09-18 05:46 & 99.00 & 99.00 & 19.2 & r \\

\hline
\caption{Forced PSF photometry of the fast optical transient ZTF20abtxwfx (AT2020sev) on images that we have acquired until one month after its first detection. Data were obtained with P48+ZTF, P200+WIRC, LCO 1-m Sinistro, and P200+WaSP.}\\
\label{tab:ZTF20abtxwfx}
\end{longtable}

\begin{longtable}{cccccc}
    \hline
    \hline
    Instrument & Time (UT) & Mag (AB) & Magerr (AB) & Lim mag (AB) & Filter \\
    \hline
ZTF & 2020-10-29 07:40 & 99.00 & 99.00 & 19.4 & g \\
ZTF & 2020-10-29 08:19 & 99.00 & 99.00 & 19.5 & r \\
ZTF & 2020-10-30 08:08 & 99.00 & 99.00 & 19.2 & g \\
ZTF & 2020-10-30 09:16 & 99.00 & 99.00 & 19.3 & r \\
ZTF & 2020-11-03 09:36 & 99.00 & 99.00 & 19.4 & g \\
ZTF & 2020-11-03 09:44 & 99.00 & 99.00 & 19.1 & g \\
ZTF & 2020-11-04 08:47 & 19.45 & 0.13 & 19.5 & g \\
ZTF & 2020-11-04 10:27 & 19.23 & 0.11 & 19.5 & r \\
ZTF & 2020-11-05 06:41 & 20.04 & 0.16 & 19.8 & g \\
ZTF & 2020-11-05 06:42 & 20.19 & 0.22 & 19.8 & g \\
ZTF & 2020-11-05 08:13 & 19.95 & 0.16 & 19.8 & r \\
ZTF & 2020-11-05 08:14 & 20.32 & 0.22 & 19.7 & r \\
GIT & 2020-11-06 19:44 & 20.97 & 0.13 & 99.00 & r \\
Sinistro & 2020-11-07 10:55 & 21.13 & 0.12 & 99.0 & g \\
Sinistro & 2020-11-07 11:01 & 20.83 & 0.15 & 99.0 & r \\
Sinistro & 2020-11-07 11:06 & 20.56 & 0.18 & 99.0 & i \\
GIT & 2020-11-07 18:16 & 21.06 & 0.04 & 99.00 & r \\
IO:O & 2020-11-07 22:56 & 21.23 & 0.08 & 23.5 & r \\
Sinistro & 2020-11-08 09:34 & 99.00 & 99.00 & 21.2 & g \\
Sinistro & 2020-11-08 09:40 & 99.00 & 99.00 & 20.8 & r \\
Sinistro & 2020-11-08 09:46 & 99.00 & 99.00 & 20.7 & i \\
GIT & 2020-11-09 15:29 & 21.91 & 0.05 & 99.00 & r \\
ZTF & 2020-11-11 07:28 & 99.00 & 99.00 & 20.1 & r \\
ZTF & 2020-11-11 08:12 & 99.00 & 99.00 & 20.4 & g \\
ZTF & 2020-11-12 07:08 & 99.00 & 99.00 & 20.8 & g \\
ZTF & 2020-11-12 07:09 & 99.00 & 99.00 & 20.7 & g \\
ZTF & 2020-11-12 07:42 & 99.00 & 99.00 & 20.7 & r \\
ZTF & 2020-11-12 07:44 & 99.00 & 99.00 & 20.7 & r \\
ZTF & 2020-11-13 07:25 & 21.30 & 0.27 & 20.6 & r \\
ZTF & 2020-11-13 09:14 & 99.00 & 99.00 & 20.6 & g \\
ZTF & 2020-11-14 07:36 & 99.00 & 99.00 & 20.2 & r \\
ZTF & 2020-11-14 07:38 & 99.00 & 99.00 & 20.2 & r \\
ZTF & 2020-11-14 08:42 & 99.00 & 99.00 & 20.0 & g \\
ZTF & 2020-11-15 06:25 & 99.00 & 99.00 & 20.7 & g \\
ZTF & 2020-11-15 06:59 & 99.00 & 99.00 & 20.4 & r \\
ZTF & 2020-11-16 06:48 & 99.00 & 99.00 & 20.5 & r \\
ZTF & 2020-11-16 06:50 & 99.00 & 99.00 & 20.4 & r \\
ZTF & 2020-11-16 07:42 & 99.00 & 99.00 & 20.5 & g \\
ZTF & 2020-11-16 07:43 & 99.00 & 99.00 & 20.5 & g \\
ZTF & 2020-11-17 05:42 & 99.00 & 99.00 & 20.6 & g \\
ZTF & 2020-11-17 06:31 & 99.00 & 99.00 & 20.3 & r \\
ZTF & 2020-11-18 06:23 & 99.00 & 99.00 & 20.6 & g \\
ZTF & 2020-11-18 08:53 & 99.00 & 99.00 & 20.5 & r \\
LRIS & 2021-01-11 06:57 & 99.0 & 99.0 & 25.6 & r \\
LRIS & 2021-01-11 06:57 & 99.0 & 99.0 & 25.5 & g \\
\hline
\caption{Forced PSF photometry of the fast optical transient ZTF20acozryr (AT2020yxz). One measurement, acquired on 2020-11-13 07:25 UT, was removed because an obviously spurious excess of flux was present at the transient location. Data were obtained with P48+ZTF, LCO 1-m Sinistro, GIT, and LT+IO:O.}
\label{tab:ZTF20acozryr}\\
\end{longtable}

\begin{longtable}{cccccc}
    \hline
    \hline
    Instrument & Time (UT) & Mag (AB) & Magerr (AB) &  Lim mag (AB) & Filter \\
    \hline
ZTF & 2021-02-03 03:10 & 99.00 & 99.00 & 20.4 & g \\
ZTF & 2021-02-03 03:11 & 99.00 & 99.00 & 20.4 & g \\
ZTF & 2021-02-03 03:37 & 99.00 & 99.00 & 19.9 & r \\
ZTF & 2021-02-03 03:55 & 99.00 & 99.00 & 20.0 & r \\
ZTF & 2021-02-03 04:19 & 99.00 & 99.00 & 19.6 & r \\
ZTF & 2021-02-03 04:39 & 99.00 & 99.00 & 20.4 & g \\
ZTF & 2021-02-04 03:16 & 99.00 & 99.00 & 18.9 & g \\
ZTF & 2021-02-04 03:17 & 99.00 & 99.00 & 19.1 & g \\
ZTF & 2021-02-04 03:17 & 99.00 & 99.00 & 19.2 & g \\
ZTF & 2021-02-04 04:17 & 99.00 & 99.00 & 19.1 & r \\
ZTF & 2021-02-04 04:47 & 99.00 & 99.00 & 19.4 & r \\
ZTF & 2021-02-04 04:48 & 99.00 & 99.00 & 19.8 & r \\
ZTF & 2021-02-04 05:42 & 99.00 & 99.00 & 17.7 & i \\
ZTF & 2021-02-05 03:11 & 99.00 & 99.00 & 20.3 & g \\
ZTF & 2021-02-05 03:57 & 99.00 & 99.00 & 19.8 & r \\
ZTF & 2021-02-05 04:10 & 99.00 & 99.00 & 19.9 & r \\
ZTF & 2021-02-05 04:53 & 99.00 & 99.00 & 20.0 & g \\
ZTF & 2021-02-06 03:11 & 20.33 & 0.13 & 20.4 & g \\
ZTF & 2021-02-06 03:14 & 20.19 & 0.12 & 20.4 & g \\
ZTF & 2021-02-06 04:16 & 20.03 & 0.13 & 20.1 & r \\
ZTF & 2021-02-06 04:48 & 19.97 & 0.12 & 20.2 & r \\
ZTF & 2021-02-07 03:25 & 21.29 & 0.31 & 20.5 & g \\
ZTF & 2021-02-07 03:25 & 99.00 & 99.00 & 20.5 & g \\
ZTF & 2021-02-07 03:58 & 20.86 & 0.25 & 20.2 & r \\
ZTF & 2021-02-07 04:12 & 21.03 & 0.27 & 20.3 & r \\
GIT & 2021-02-07 18:22 & 21.31 & 0.04 & 99.0 & r \\
ZTF & 2021-02-08 03:15 & 99.00 & 99.00 & 19.5 & i \\
ZTF & 2021-02-08 03:47 & 99.00 & 99.00 & 20.1 & r \\
ZTF & 2021-02-08 04:15 & 99.00 & 99.00 & 20.4 & g \\
GIT & 2021-02-08 15:14 & 22.22 & 0.06 & 23.05 & r \\
ZTF & 2021-02-09 03:25 & 99.00 & 99.00 & 20.3 & g \\
ZTF & 2021-02-09 03:25 & 99.00 & 99.00 & 20.4 & g \\
ZTF & 2021-02-09 03:58 & 99.00 & 99.00 & 20.1 & r \\
ZTF & 2021-02-09 03:59 & 99.00 & 99.00 & 20.2 & r \\
ZTF & 2021-02-10 03:15 & 99.00 & 99.00 & 20.3 & r \\
ZTF & 2021-02-10 03:25 & 99.00 & 99.00 & 20.2 & r \\
Sinistro & 2021-02-10 03:38 & 99.00 & 99.00 & 21.6 & i \\
ZTF & 2021-02-10 03:45 & 99.00 & 99.00 & 20.5 & g \\
ZTF & 2021-02-10 03:47 & 99.00 & 99.00 & 20.5 & g \\
Sinistro & 2021-02-10 03:54 & 99.00 & 99.00 & 22.6 & g \\
ZTF & 2021-02-11 03:25 & 99.00 & 99.00 & 20.2 & g \\
ZTF & 2021-02-11 03:54 & 99.00 & 99.00 & 20.2 & r \\
ZTF & 2021-02-11 04:03 & 99.00 & 99.00 & 20.3 & r \\
ZTF & 2021-02-11 04:46 & 99.00 & 99.00 & 19.6 & i \\
ZTF & 2021-02-11 06:19 & 99.00 & 99.00 & 19.9 & g \\
ZTF & 2021-02-12 03:12 & 99.00 & 99.00 & 20.1 & r \\
ZTF & 2021-02-15 03:41 & 99.00 & 99.00 & 19.8 & r \\
ZTF & 2021-02-15 04:20 & 99.00 & 99.00 & 20.1 & g \\
ZTF & 2021-02-15 04:21 & 99.00 & 99.00 & 20.0 & g \\
ZTF & 2021-02-15 05:23 & 99.00 & 99.00 & 18.6 & r \\
ZTF & 2021-02-18 03:13 & 99.00 & 99.00 & 19.5 & g \\
ZTF & 2021-02-18 03:14 & 99.00 & 99.00 & 19.5 & g \\
ZTF & 2021-02-20 03:17 & 99.00 & 99.00 & 19.8 & r \\
ZTF & 2021-02-20 03:44 & 99.00 & 99.00 & 19.9 & g \\
ZTF & 2021-02-21 03:20 & 99.00 & 99.00 & 19.5 & g \\
ZTF & 2021-02-21 04:12 & 99.00 & 99.00 & 19.0 & i \\
ZTF & 2021-02-21 05:20 & 99.00 & 99.00 & 19.1 & r \\
ZTF & 2021-02-22 03:19 & 99.00 & 99.00 & 19.6 & r \\
ZTF & 2021-02-22 04:43 & 99.00 & 99.00 & 19.3 & g \\
ZTF & 2021-02-23 03:50 & 99.00 & 99.00 & 19.7 & g \\
ZTF & 2021-02-23 05:45 & 99.00 & 99.00 & 19.2 & r \\
    \hline
\caption{Forced PSF photometry of the fast optical transient ZTF21aahifke (AT2021clk). Data  were  obtained  with  P48+ZTF, GIT, and LCO 1-m Sinistro.}
\label{tab:ZTF21aahifke}\\
\end{longtable}

\begin{longtable}{cccccc}
    \hline
    \hline
    Instrument & Time (UT) & Mag (AB) & Magerr (AB) &  Lim mag (AB) & Filter \\
    \hline
ZTF & 2021-02-03 05:58 & 99.00 & 99.00 & 20.2 & r \\
ZTF & 2021-02-03 05:59 & 99.00 & 99.00 & 20.4 & r \\
ZTF & 2021-02-03 06:26 & 99.00 & 99.00 & 20.7 & g \\
ZTF & 2021-02-03 06:28 & 99.00 & 99.00 & 20.7 & g \\
ZTF & 2021-02-03 07:58 & 99.00 & 99.00 & 19.5 & i \\
ZTF & 2021-02-04 05:14 & 99.00 & 99.00 & 18.3 & g \\
ZTF & 2021-02-04 07:07 & 17.16 & 0.03 & 19.2 & r \\
ZTF & 2021-02-05 05:09 & 19.31 & 0.06 & 20.2 & r \\
ZTF & 2021-02-05 06:07 & 19.75 & 0.08 & 20.5 & g \\
ZTF & 2021-02-05 06:08 & 19.67 & 0.08 & 20.4 & g \\
ZTF & 2021-02-05 07:52 & 19.41 & 0.07 & 20.2 & r \\
ZTF & 2021-02-06 05:26 & 19.99 & 0.08 & 20.6 & r \\
ZTF & 2021-02-06 05:35 & 19.71 & 0.10 & 20.1 & i \\
ZTF & 2021-02-06 09:27 & 20.62 & 0.20 & 20.2 & g \\
ZTF & 2021-02-07 05:06 & 20.56 & 0.14 & 20.6 & r \\
ZTF & 2021-02-07 06:52 & 21.01 & 0.18 & 20.8 & g \\
ZTF & 2021-02-08 05:30 & 99.00 & 99.00 & 19.8 & i \\
ZTF & 2021-02-09 05:32 & 99.00 & 99.00 & 20.5 & g \\
ZTF & 2021-02-09 06:05 & 21.35 & 0.25 & 20.7 & r \\
ZTF & 2021-02-09 07:43 & 21.09 & 0.36 & 20.0 & i \\
ZTF & 2021-02-11 05:04 & 99.00 & 99.00 & 20.6 & r \\
ZTF & 2021-02-11 06:36 & 99.00 & 99.00 & 20.8 & g \\
ZTF & 2021-02-12 07:22 & 99.00 & 99.00 & 19.8 & i \\
ZTF & 2021-02-15 04:35 & 99.00 & 99.00 & 20.2 & g \\
ZTF & 2021-02-15 06:00 & 99.00 & 99.00 & 20.3 & r \\
ZTF & 2021-02-15 07:44 & 99.00 & 99.00 & 19.7 & i \\
ZTF & 2021-02-18 05:07 & 99.00 & 99.00 & 19.8 & r \\
ZTF & 2021-02-18 06:38 & 99.00 & 99.00 & 19.9 & g \\
ZTF & 2021-02-18 07:09 & 99.00 & 99.00 & 19.3 & i \\
    \hline
\caption{Forced PSF photometry of the fast optical transient ZTF21aagwbjr (AT2021buv), the optical counterpart to GRB~210204A, on images obtained with P48+ZTF. Full multi-wavelength data analysis will be presented by Kumar et al. (in preparation).}
\label{tab:ZTF21aagwbjr}\\
\end{longtable}

\begin{longtable}{cccccc}
    \hline
    \hline
    Instrument & Time (UT) & Mag (AB) & Magerr (AB) &  Lim mag (AB) & Filter \\
    \hline
ZTF & 2021-03-07 08:22 & 99.00 & 99.00 & 19.9 & i \\
ZTF & 2021-03-07 09:27 & 99.00 & 99.00 & 20.5 & g \\
ZTF & 2021-03-07 10:17 & 99.00 & 99.00 & 20.2 & r \\
ZTF & 2021-03-08 06:17 & 99.00 & 99.00 & 20.1 & r \\
ZTF & 2021-03-08 06:27 & 99.00 & 99.00 & 18.2 & r \\
ZTF & 2021-03-17 09:23 & 19.74 & 0.06 & 20.9 & r \\
ZTF & 2021-03-17 09:41 & 19.99 & 0.07 & 20.9 & g \\
ZTF & 2021-03-18 07:54 & 20.19 & 0.15 & 20.4 & i \\
ZTF & 2021-03-18 09:22 & 20.12 & 0.10 & 20.6 & r \\
ZTF & 2021-03-18 11:56 & 20.36 & 0.12 & 20.6 & g \\
ZTF & 2021-03-19 09:19 & 20.36 & 0.10 & 20.8 & r \\
ZTF & 2021-03-19 10:18 & 20.59 & 0.12 & 20.8 & g \\
    \hline
\caption{Forced PSF photometry of the fast optical transient ZTF21aapkbav (AT2021gca), a Type II supernova, on images obtained with P48+ZTF.}
\label{tab:ZTF21aapkbav}\\
\end{longtable}

\end{appendix}

\end{document}